\documentclass[8pt]{article}
\usepackage{graphicx}
\usepackage{pgfplots}
\usepackage{newlfont}
\usepackage[colorlinks,linkcolor=blue,citecolor=blue,anchorcolor=blue,bookmarksopen,pdfpagetransition={Wipe}]{hyperref}
\usepackage{amsmath,amsthm,amssymb}
\usepackage{subfigure}
\usepackage{amsmath}
\usepackage{booktabs}
\usepackage{fourier}
\usepackage{array}
\usepackage{makecell}
\usepackage{algorithm}
\usepackage{algpseudocode}

\newtheorem{thm}{Theorem}[section]

\newtheorem{rem}[thm]{Remark}
\numberwithin{equation}{section}

\newcommand{\h}{\mathcal{H}}

\begin{document}
\title{\textbf{2D Hybrid chaos map for image security transform based on framelet and cellular automata  }}
\author{Y. Khedmati$^a$\footnote{khedmati.y@uma.ac.ir, khedmatiy.y@gmail.com},
R. Parvaz$^a$\footnote{Corresponding author: rparvaz@uma.ac.ir}\
, Y. Behroo$^a$\footnote{yousef.behroo@uma.ac.ir}}
\date{}
\maketitle
\begin{center}
$^a$Department of Mathematics, University of Mohaghegh Ardabili,
56199-11367 Ardabil, Iran.\\
\end{center}
\begin{abstract}
In this paper, we provide some safe ways to transfer images securely by using cryptography and steganography methods.
In order to enhance the security of the image transmission,
we introduce a new type of uniformly distributed 2D-hybrid chaos map based on Logistic, Sine and Tent maps, and use the cellular automata and discrete framelet transform in the proposed algorithms and also
mix the position of the image pixels by apply kinds of shifts.
To show that the proposed algorithms are able to resist various attacks, different types of
 simulation results and security analysis are used.
\indent
\end{abstract}
\vskip.3cm

\section{Introduction and Preliminaries}
\indent \hskip.65cm
In the last decade, with the development of the social networks
as WatsApp, Facebook, Instagram and etc.,
the importance of information security has increased.
Since images play an important role in the social networks,
creating a safe environment for transferring images is important.
Encryption is one of the methods for
safe transfer of image.
Chaotic system is one of the most important tools in cryptography.
Early studies on chaotic systems for encryption has been given by \cite{1},
and in recent years, this method has been used and developed in many papers
\cite{2,3,4,5}. Another tool used in cryptography is the cellular automata.
This tool is introduced by Von Neumann \cite{6,7} and is used in various articles for
cryptography \cite{8,9,10}.
Steganography is the other method for image security
transform.
This method is used
in many papers for image security transform \cite{45,46}.
The spatial domain and the transform domain are
common methods in the steganography \cite{44}.
The Least Significant Bit (LSB) is considered as the main technique in spatial domain \cite{48,49,50}.
Another tool that is used in this article is the framelet transform.
In 1952, the concept of frames for Hilbert spaces was defined by Duffin and Schaeffer \cite{11}.
Although the details of the framelet transform can be found in the \cite{12}, in
the continue of this section,
some basic relationships to this discussion are given.
This type of transition has applications in various area of image processing such as
deblurring, denoising \cite{13,14}.
In the present paper, in the first step, a 2D hybrid chaotic system has been introduced, then
this system and the cellular automata method have been used in encryption algorithm. In the last step,
an image hidden method based on the hybrid chaotic system and the framelet transform is introduced.
The following subsections provide a brief overview of the basic principles.

\subsection{Hybrid Chaos Map}
The basic chaotic systems have some weaknesses.
The nonuniform distribution and area limit of chaos behavior are
the most important of these weaknesses.
Combining these types of systems is one of
 the best ways to minimize these weaknesses.
The first attempt to combine the two systems can be referred to Logistic Tent system (LTS) \cite{15}.
Subsequently, various methods have been introduced in the articles for hybrid systems,
as Tent-Sine system \cite{16} and 2D Sine Logistic modulation map (2D SLMM) \cite{17}.
\subsection{Frame Theory}
Let $\h$ be a separable Hilbert space. We call a sequence $F=\{f_i\}_{i\in I} \subseteq \h$
a frame for $\h,$ if there exist two constant $A_F, B_F> 0$ such that
\begin{eqnarray}\label{abc}
A_F \|f\|^2\leq\sum_{i\in I}|\langle f,f_i\rangle |^2 \leq B_F \|f\|^2,\quad f\in \h.
\end{eqnarray}
If in (\ref{abc}), $A_F=B_F=1$ we say that $F=\{f_i\}_{i\in I}$ is a Parseval
frame for $\h$. Let $F=\{f_i\}_{i\in I}$ be a frame for $\h,$ then the operator
\begin{eqnarray*}
T_F:l_2(I) \rightarrow \h,\quad T_F(\{c_i\}_{i\in I})=\sum_{i\in I} c_if_i,
\end{eqnarray*}
is well define and onto, also its adjoint is
\begin{eqnarray*}
T^*_F:\h\rightarrow l_2(I) ,\quad T_F^* f=\{\langle f,f_i \rangle\}_{i\in I}.
\end{eqnarray*}
The operators $T_F$ and $T^*_F$ are called the synthesis and analysis operators of
the frame $F.$ By composing $T_F$ and $T^*_F,$ we obtain the frame operator
\begin{align*}
S_F:\h\rightarrow\h,\quad S_Ff=\sum_{i=1}^{\infty}\langle f,f_i\rangle f_i,\quad f\in \h.
\end{align*}
The operator $S_F$ is a positive, self-adjoint invertible operator on $\h$ with
$A_F.I_\h\leq S_F\leq B_F.I_\h,$ where $I$ is the identity operator on $\h$ and $A_F$ and $B_F$
are the lower and upper frame bounds, respectively \cite{12}. For Parseval frame
$F=\{f_i\}_{i\in I},$ we have $S_F=I_\h.$
\begin{thm}\cite{12}
Let $F=\{f_i\}_{i\in I}$ be a frame for $\h$ with frame operator $S_F.$ Then
\begin{align*}
f=\sum_{i=1}^{\infty}\langle f,S_F^{-1}f_i\rangle f_i,\quad f\in \h,
\end{align*}
and
\begin{align*}
f=\sum_{i=1}^{\infty}\langle f,f_i\rangle S_F^{-1}f_i,\quad f\in \h.
\end{align*}
\end{thm}
In the proposed algorithm for image transform, we apply particular Parseval framelet
systems in $\h=L_2(\mathbb R)$ that were constructed from $B$-spline whose refinement
 mask is $h_0=1/4[1,2,1],$ with two corresponding framelet masks $h_1=\sqrt{2}/4[1,0,-1]$
 and $h_0=1/4[-1,2,-1]$.
For the details of the image transform by using framelet,
 the reader can refer to \cite{18,19}.

\subsection{Cellular automata}
\begin{figure}
\centering
\includegraphics*[width=.80\textwidth]{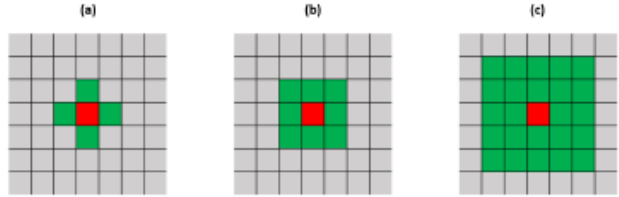}\\
\vspace{-2cm}
\emph{\caption{
(a) Von Neumann neighborhood, (b) Moore neighborhood, (c) extended Moore neighborhood.
}\label{f1}}
\end{figure}

The cellular automata can be considered as a mathematical model for the discrete
dynamical systems which composed of a number of cells.
These cells together build a network, so that they are updated according to the
special rules.
In the following of this paper, the general principles used in the paper are
explained
and more details about this area can be found in \cite{20,21}.
In the proposed algorithms, we use the combined cellular automata.
The combined cellular automata performance is like the common cellular automata, with the difference
that the combined cellular automata can assign different rule number for each cell.
The different dimensions can be classified for the cellular
automata as one-dimensional, two-dimensional and three-dimensional.
In this paper, one-dimensional cellular automata has been used.
Also for using the cellular automata, there is a need for boundary conditions.
For the cellular automata, we can consider periodic, reflective, and fixed value boundaries as
boundary conditions. In the proposed algorithms, periodic boundary condition as follows has been used.
\begin{align}
S^{t}_{i,j}=S^t_{u,v}\Leftrightarrow i\equiv u(mod~n) and j\equiv v(mod~m),
\end{align}

where $i,j,u$ and $v$ denote cell coordinates in two-dimensional space of the size
$n\times m$ and $S$ is cell state at the time step $t$.
Also, in this method, different neighbors are considered as the von Neumann
neighborhood, the Moore neighborhood and the extended Moore neighborhood
(see Fig. \ref{f1}).
In the proposed methods, the von Neumann neighborhood is used.
The cellular automata are divided into two categories: reversible and irreversible.
In the proposed key generation algorithm, irreversible cellular automata is used,
and in encryption algorithm,
reversible cellular automata is used.
\begin{figure}
\centering
\subfigure{
\includegraphics*[width=.90\textwidth]{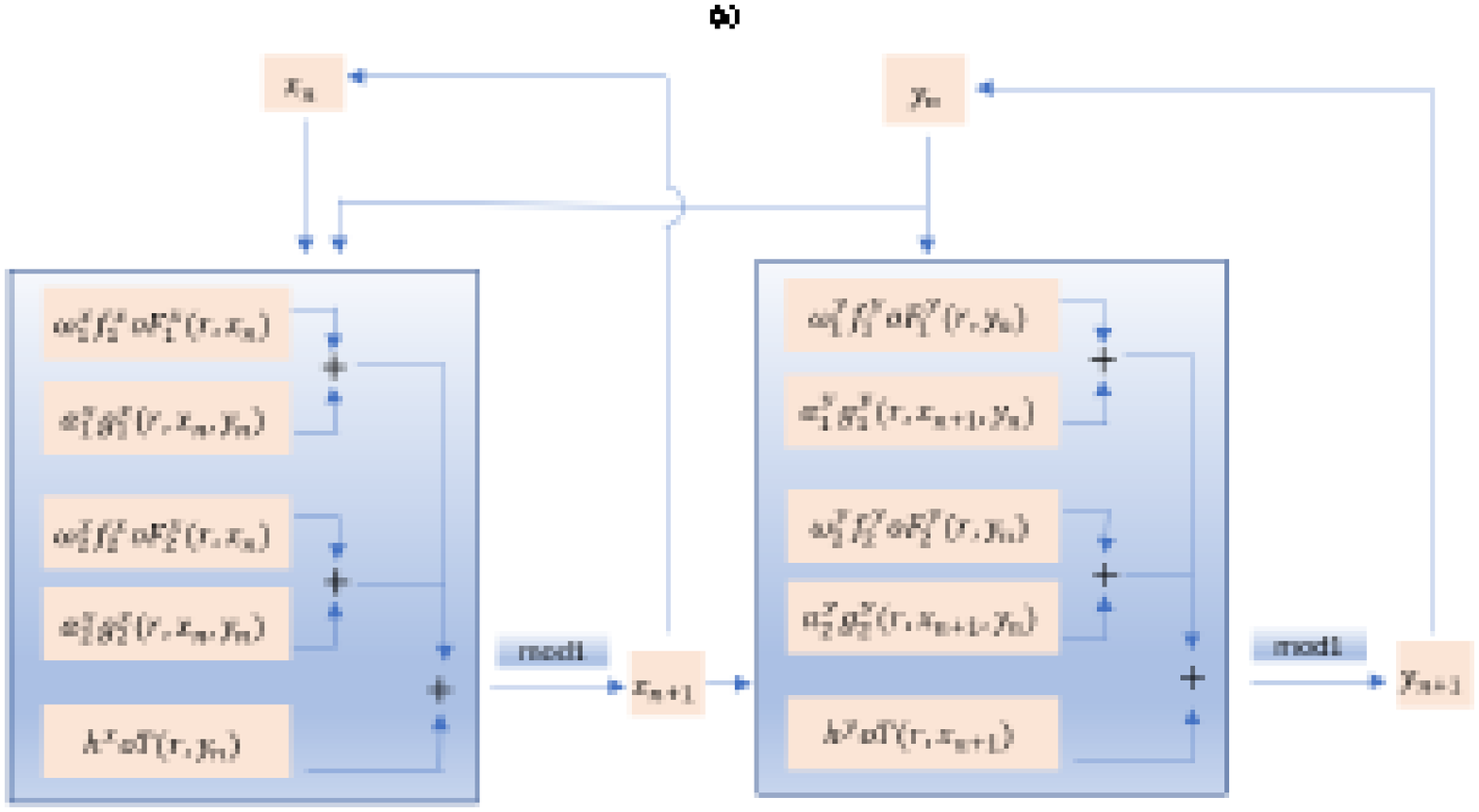}}\\
\vspace{-1.25cm}
\subfigure{
\includegraphics*[width=.90\textwidth]{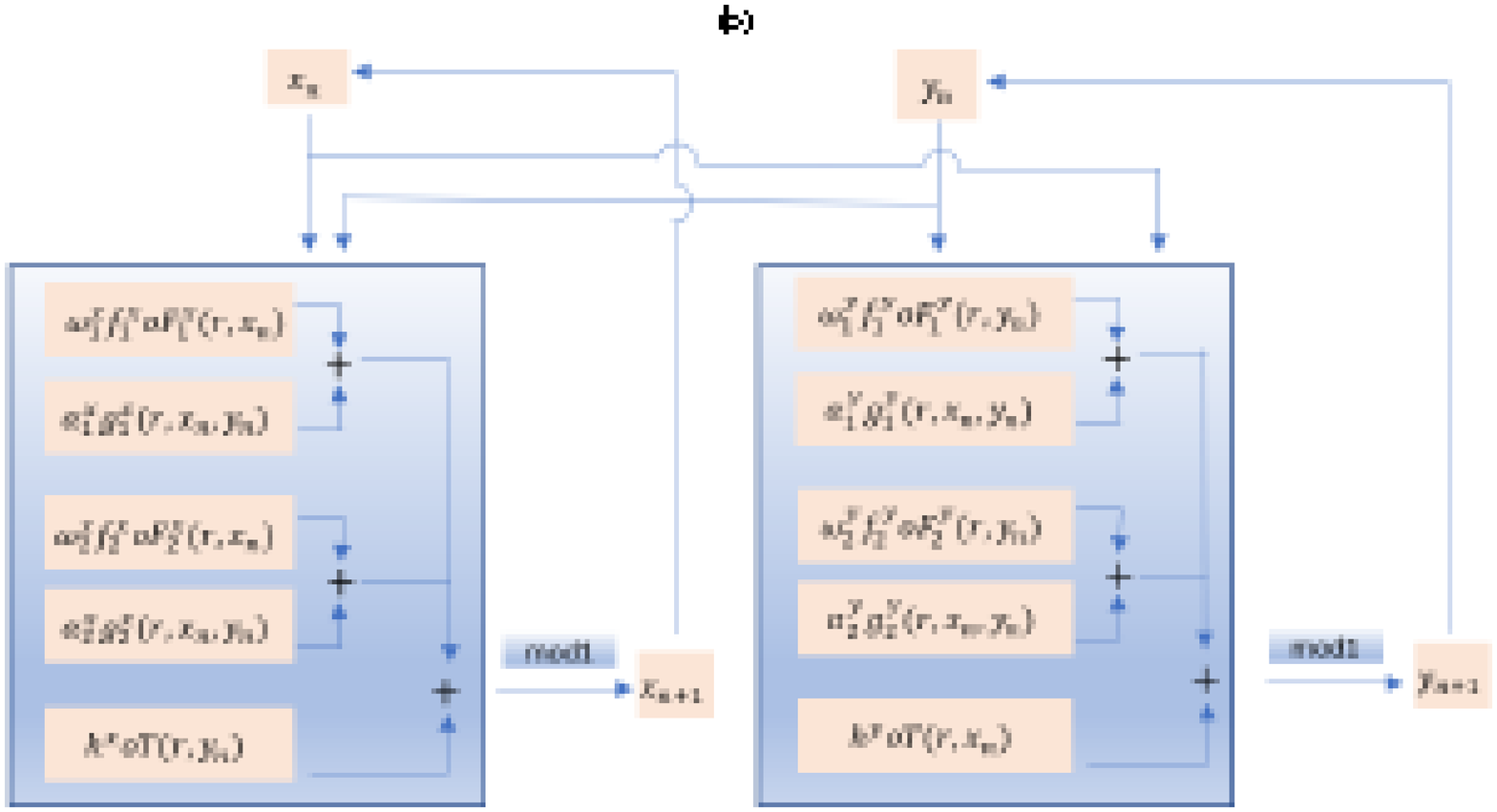}}\\
\vspace{-1.25cm}
\emph{\caption{
Proposed hybrid chaotic systems.
}\label{f2}}
\end{figure}
\section{Two dimensional hybrid chaotic system}
\indent \hskip .65cm
In this section, we describe two-dimensional combination chaotic system based on Logistic, Sine and Tent maps.
One-dimensional
Logistic,  Sine and Tent maps are defined as follows \cite{22}:
\begin{align}
&x_{i+1}=L(r,x_i):=rx_i(1-x_i),\\
&x_{i+1}=S(r,x_i):=r\sin(\pi x_i)/4,
\end{align}
\begin{align}
x_{i+1}=T(r,x_i):=
\left\{%
\begin{array}{ll}
rx_i/2,&\textnormal{when}~x_i<0.5,\\
\\
r(1-x_i)/2,&\textnormal{when}~x_i\geq0.5.\\
\end{array}%
\right.
\end{align}

One of the two-dimensional chaos map is the Logistic map, which has
the following form \cite{23}
\begin{align}
&x_{i+1}=r(3y_i+1)x_i(1-x_i),\nonumber\\
&y_{i+1}=r(3x_{i+1}+1)y_{i}(1-y_i).
\end{align}
Also in \cite{17}, 2D Sine Logistic modulation map (2D SLMM) is introduced as follows
\begin{align}
&x_{i+1}=\alpha(\sin(\pi y_i)+\beta)x_i(1-x_i),\nonumber\\
&y_{i+1}=\alpha(\sin(\pi x_{i+1})+\beta)y_i(1-y_i).
\end{align}
Each of the above maps has weaknesses that we will review.
Flat histogram and uniform distribution are two most important
features of the ideal chaos map for encryption process.
Nonuniform distribution over output series can cause weakness
in the decryption attacks as statistical attack.
The most important and basic chaos maps as Logistic or Tent are not well enough
to estimate these conditions.
However, a good combination and transformation of these maps can satisfy these conditions.
In the proposed hybrid chaos map, combining, transferring and adding weights are used.
In the following, two combinational algorithms are used to obtain the hybrid chaos maps.
The general outline of the combinational algorithms are presented in Fig. \ref{f2}.
In the first step of the combinational algorithms \textbf{(a)} and \textbf{(b)},
$x_n$ and $y_n$ are used as inputs.
Regarding the combinational algorithms (see Fig. \ref{f2}), it can be seen that
the difference between the two algorithms is in the input of the second step.
In the first algorithm, the output value of the first step (i.e. $x_{n+1}$)
and $y_n$ are used as input in the second step, but in the
second algorithm, $x_n$ and $y_n$ are used as input values for the next step.
The mathematical formulae for the algorithms can be expressed as follows
\begin{align}
x_{i+1}:=
\left\{%
\begin{array}{ll}
\omega^x_1f^x_1 \circ F^x_1(r,x_i)+\alpha^x_1g^x_1(r,x_i,y_i)\\
~~~~~~~~~~~~+h_1^x(\frac{(\beta^x_1-r)x_i}{2})~mod~1,~~\textnormal{when}~y_i<0.5,\\
\\
\omega^x_2f^x_2 \circ F^x_2(r,x_i)+\alpha^x_2g^x_2(r,x_i,y_i)\\
~~~~~~~~~~~~+h_2^x(\frac{(\beta^x_2-r)(1-x_i)}{2})~mod~1,~~\textnormal{when}~y_i\geq0.5,\\
\end{array}%
\right.
\end{align}

\begin{align}
y_{i+1}:=
\left\{%
\begin{array}{ll}
\omega^y_1f^y_1 \circ F^y_1(r,y_i)+\alpha^y_1g^y_1(r,\zeta,y_i)\\
~~~~~~~~~~~~+h_1^y(\frac{(\beta^y_1-r)\zeta}{2})~mod~1,~~\textnormal{when}~\zeta<0.5,\\
\\
\omega^y_2f^y_2 \circ F^y_2(r,y_i)+\alpha^y_2g^y_2(r,\zeta,y_i)\\
~~~~~~~~~~~~+h_2^y(\frac{(\beta^y_2-r)(1-\zeta)}{2})~mod~1,~~\textnormal{when}~\zeta\geq0.5,\\
\end{array}%
\right.
\end{align}
where
\begin{align}
\zeta:=
\left\{%
\begin{array}{ll}
x_i,~~~~\textnormal{for Algorithm (a)},\\
\\
x_{i+1},~~~~\textnormal{for Algorithm (b)}.\\
\end{array}%
\right.
\end{align}
In the above formula, $\omega^k_l,\alpha^k_l,\beta_l^k~(k=x,y\,\&\,l=1,2)$
are considered as weights,
$f^{k}_l,h^{k}_l$ $(k=x,y\,\&\,l=1,2)$
are used as combination maps and $g^{k}_l~(k=x,y\,\&\,l=1,2)$ act as
transfer maps.
Also $F^k_l(k=x,y\,\&\,l=1,2)$ are chosen as an arbitrary map between
Logistic and Sine maps.
By choosing different values and maps for weights, combination and
transfer maps,
different chaotic systems are created. To investigate the properties of
the proposed hybrid chaos map,
we study several examples of these maps.\\

Consider the following cases: \\

\noindent\textbf{Case (i)}:  $\omega^x_1=\omega^y_1=10,\omega^x_2=\omega^y_2=20,
\alpha^x_1=\alpha^y_1=\alpha^y_2=2,\alpha^x_2=7,\zeta=x_{i+1},$\\
\indent \hskip 0.7cm$f_1^x(p)=\tan(p), f_2^x(p)=f_2^y(p)=\sin(p),f_2^y(p)=p,$\\
\indent\hskip 0.7cm $g_1^x(r,q,p)=\exp(rp)+\exp(rq),g^x_2(r,q,p)=rp+\frac{2}{7}\exp(\pi rq),$\\
\indent\hskip 0.7cm $g_1^y(r,q,p)=\tan(rq+p),g^y_2(r,q,p)=\exp(20rq),$\\
\indent\hskip 0.7cm $\beta^x_1=80,\beta^x_2=20,\beta^y_1=50,\beta^y_2=30,$\\
\indent\hskip 0.7cm $h^x_1(p)=\sin(2p),h^x_2(p)=\sin(4p),h^y_1(p)=\exp(2p),h^y_2(p)=\cos(4p),$\\
\indent\hskip 0.7cm $F_l^k~(k=x,y\,\&\,l=1,2)$ are considered as Logistic map.\\

\noindent\textbf{Case (ii)}:  $\omega^x_1=1,\omega^y_1=16,\omega^x_2=10,
\omega^y_2=20,\alpha^x_1=15,\alpha^x_2=7,\alpha^y_1=2,\alpha^y_2=14,,\zeta=x_{i+1},$\\
\indent\hskip 0.7cm $f_1^x(p)=\cos(p), f_2^x(p)=cot(p),f_2^y(p)=p,f_2^y(p)=\sin(\pi p),$\\
\indent\hskip 0.7cm $g_1^x(r,q,p)=rp+\frac{12}{15}\cos(rq),g^x_2(r,q,p)=-rp+\log(\pi rq),$\\
\indent\hskip 0.7cm $g_1^y(r,q,p)=\tan(rq+p),g^y_2(r,q,p)=\exp(20rq),$\\
\indent\hskip 0.7cm $\beta^x_1=26,\beta^x_2=2,\beta^y_1=50,\beta^y_2=30,$\\
\indent\hskip 0.7cm $h^x_1(p)=\sin(2p),h^x_2(p)=\exp(4p),h^y_1(p)=\exp(2p),h^y_2(p)=\cot(4p).$\\

\noindent\textbf{Case (iii)}:  $\omega^x_1=10,\omega^y_1=20,\omega^x_2=5,
\omega^y_2=30,\alpha^x_1=2,\alpha^x_2=7,\alpha^y_1=2,\alpha^y_2=4,,\zeta=x_{i},$\\
\indent\hskip 0.7cm $f_1^x(p)=p, f_2^x(p)=coth(p),f_2^y(p)=sin(p),f_2^y(p)=p,$\\
\indent\hskip 0.7cm $g_1^x(r,q,p)=\sin(rp)+2\exp(rq),g^x_2(r,q,p)=\exp(20rq)+\sin(\pi r x),$\\
\indent\hskip 0.7cm $g_1^y(r,q,p)=p\tan(rq),g^y_2(r,q,p)=\cos(rq),$\\
\indent\hskip 0.7cm $\beta^x_1=80,\beta^x_2=20,\beta^y_1=50,\beta^y_2=3,$\\
\indent\hskip 0.7cm $h^x_1(p)=\sin(2p),h^x_2(p)=\cos(4p),h^y_1(p)=\sinh(2p),h^y_2(p)=4p.$\\

\begin{figure}
\centering
\includegraphics*[height=7cm, width=12cm]{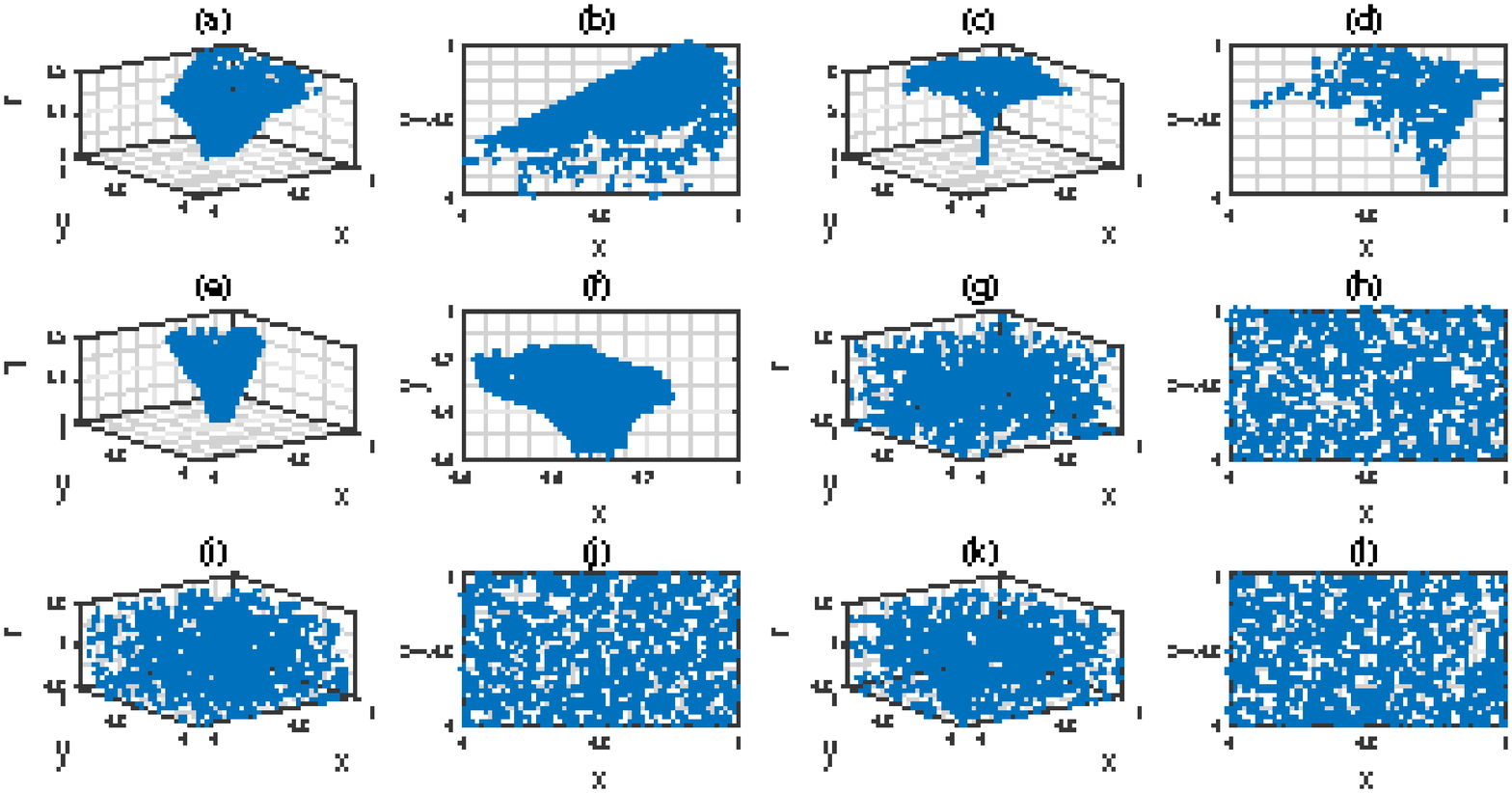}
\vspace{-0.99cm}
\emph{\caption{
 Distribution patterns: (a-b) 2D Logistic map, (c-d) 2D SLMM for
 different value of $\beta$ with $\alpha=1$,
 (e-f) 2D-SLMM for different value of $\alpha$ with $\beta=2$, (g-h)
 Case (i) with $r=1.19$, (i-j) Case (ii) with $r=1.19$
(k-l) Case (iii) with $r=1.19$.
}\label{f3}}
\end{figure}

\begin{figure}
\centering
\includegraphics*[height=7cm, width=12cm]{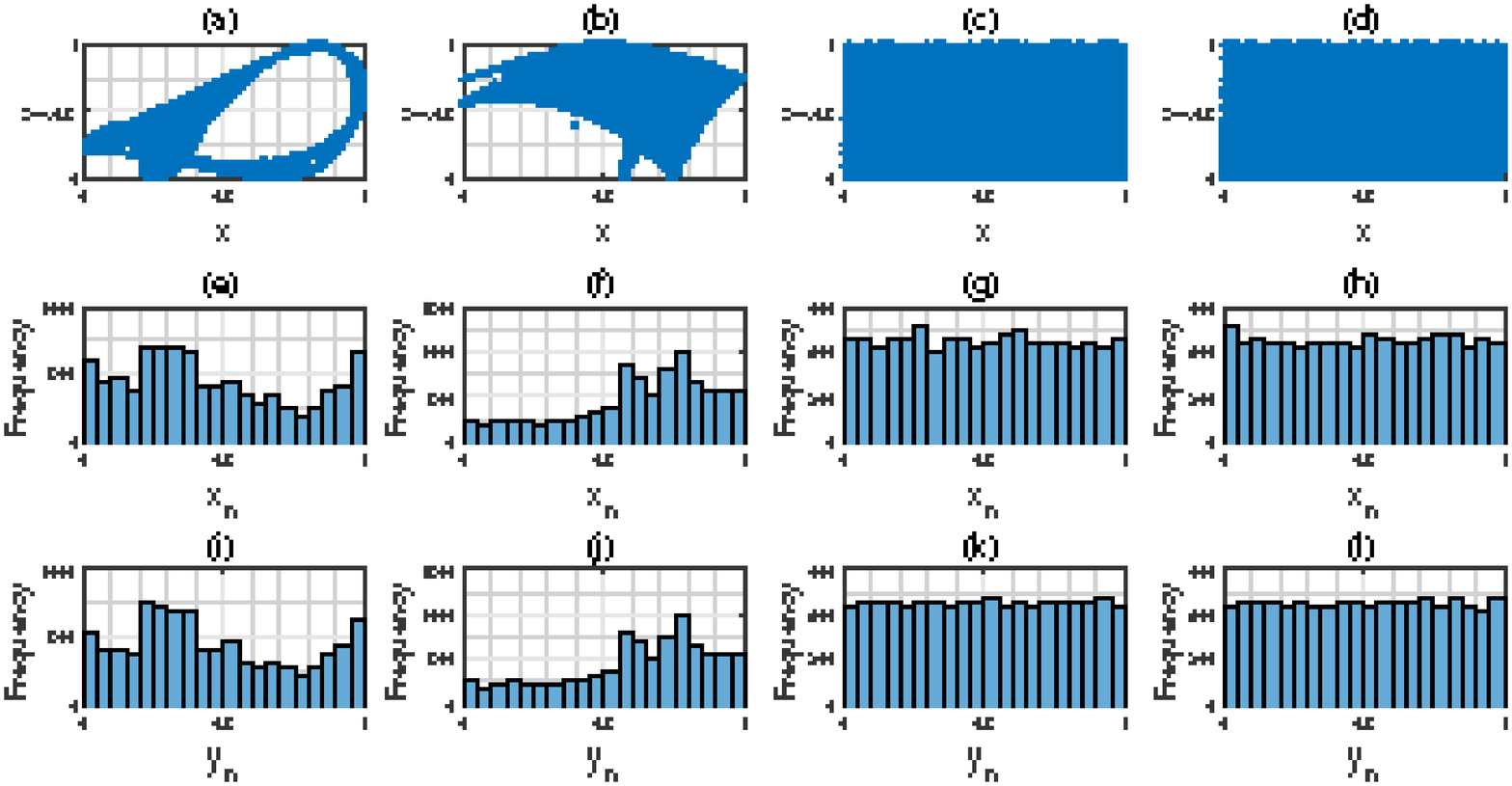}
\vspace{-0.99cm}
\emph{\caption{
Distribution and histogram plots: (a-e-i) 2D-Logistic map with 1.19,
(b-f-j) 2D-SLMM with $\alpha=1,\beta=3$, (c-g-k) Case (i) with $r=1.19$,
(d-h-l) Case (iii) with $r=1.19$.
}\label{f4}}
\end{figure}

\begin{figure}
\centering
\includegraphics*[height=3cm, width=12cm]{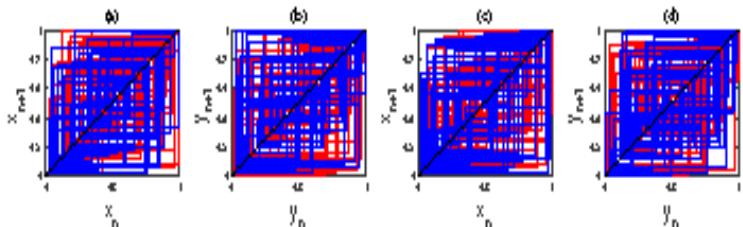}
\vspace{-0.5cm}
\emph{\caption{
Cobweb plot: (a-b) Case (i), (c-d) Case (ii).
}\label{f5}}
\end{figure}

\begin{figure}
\centering
\includegraphics*[width=.80\textwidth]{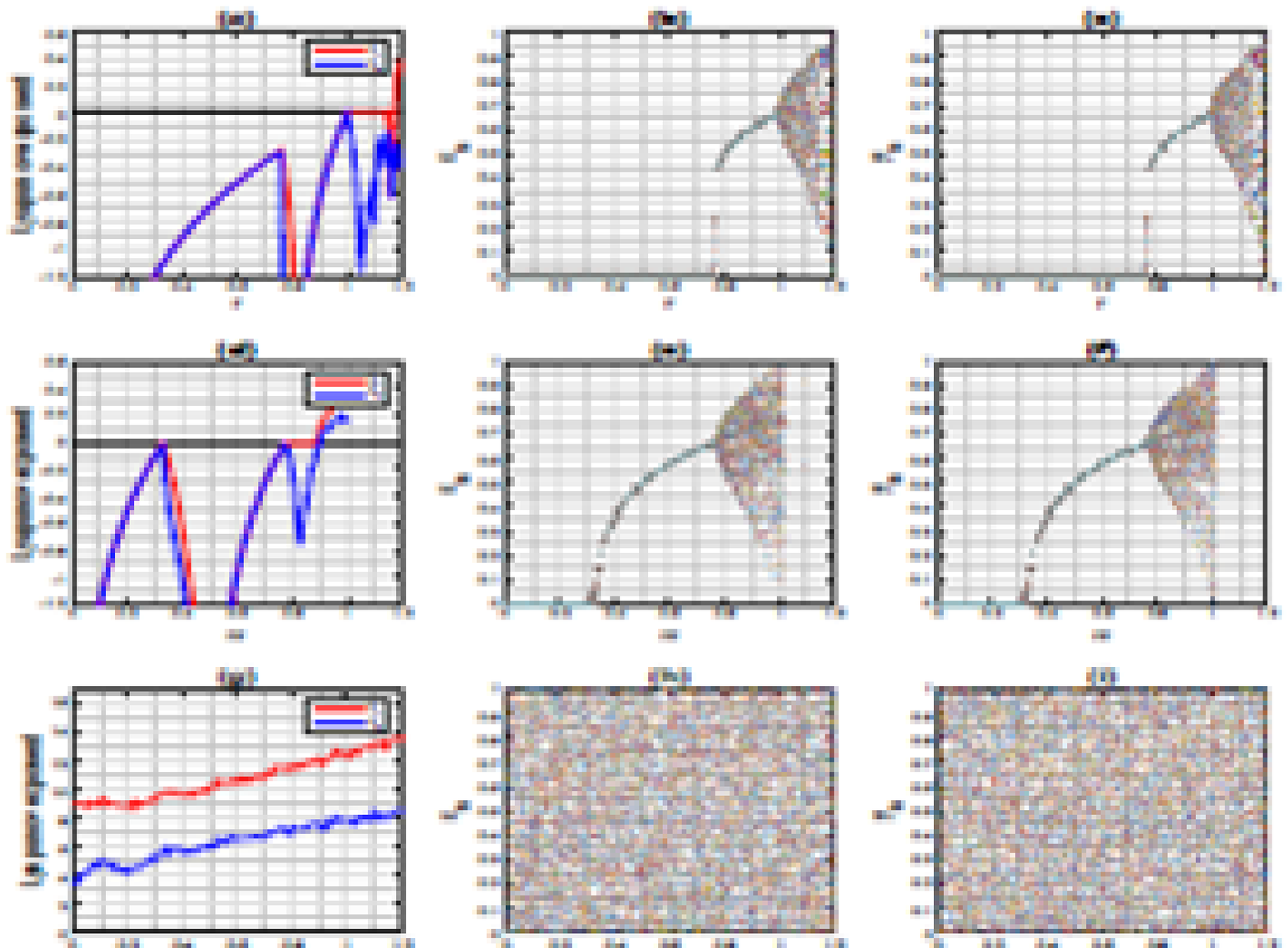}
\emph{\caption{
 Lyapunov exponent and bifurcation diagram results for
(a-c) 2D-Logistic map, (d-f) 2D-SLMM for $\beta=3$,
(g-i) proposed hybrid chaotic system case (iii).
}\label{f6} }
\end{figure}

In Case (ii) and Case (iii), $F_1^x,F_2^y$ are considered as Logistic maps and $F_2^x,F_1^y$ are considered as
Sine maps. Results for the above cases are presented in Fig.s \ref{f3}-\ref{f6}.
In the Fig. \ref{f3}, for the proposed hybrid chaos map, 2D Logistic map and 2D SLMM distribution patterns
for $100$ points
are given.
With regard to the results, it can be seen that the proposed maps have a
uniform distribution compared
to the other two maps. The histograms of output data are shown in Fig. \ref{f4}.
As seen from this figure, the histograms of 2D Logistic map and 2D SLMM are not flat, while the histograms
of the proposed maps in comparison with two maps have a flat distribution.
Also, cobweb plots in Fig. \ref{f5} show chaotic behavior for Case (i) and Case (ii).
In this figure, $100$ points generated by the chaos maps are used.
Lyapunov Exponents is one of the most important value in the study
of the behavior of the chaotic systems.
Sensitivity to initial conditions can be studied by using this value.
Different methods have been introduced in some articles to calculate
this value, for example, the reader
can see \cite{24,25,26}. In this section, the method in \cite{25} is used to
find the Lyapunov Exponent.
The relationship between Lyapunov exponent and chaotic system has been
studied in various articles \cite{27,28}.
The following sentence is stated in \cite{28}:\\
``\textit{In an $n$ dimensional dynamical system, we have $n$
Lyapunov exponents. Each $\lambda_k$ represents the divergence of $k$-volume ($k=1:$
 length,k=2: area, etc.). The sign of the Lyapunov exponents indicates the behavior of
nearby trajectories. A negative exponent indicates that neighboring trajectories
converge to the same trajectory. A positive exponent indicates that neighboring
trajectories diverge} \cite{28}''.\\
In Fig.6, $2000$ points are generated by the chaos maps are used to calculate
the Lyapunov exponents.
According to the results in Fig. \ref{f6},
it is seen that in the larger interval, in comparison with 2D Logistic
map, the proposed hybrid map
has chaotic behavior.
Another convenient tool for chaos study is bifurcation diagram.
The bifurcation diagram can be used to show chaotic attractors.
The results for bifurcation diagrams are given in Fig. \ref{f6}.
Considering the above discussions, it can be concluded
that the proposed chaos maps have good chaotic behavior such as
flat distribution, sensitivity to initial condition and
unpredictability of the output points.


 \section{{\textbf Encryption and Decryption Algorithms}}
 In this section, the details of the encryption and decryption algorithms are given.
\begin{rem}
In the reminder of this paper, $\varpi(x_0,y_0,r,n)$ denotes the matrix
$(a_{i,j})\in \mathcal{R}^{2\times n},$ where
\begin{align*}
a_{i,j}:=
\left\{%
\begin{array}{ll}
x_j,&~i=1,\\
\\
y_j,&~i=2,\\
\end{array}%
\right.
\end{align*}
\noindent and $x_j, y_j \:(j=1,2,...,n)$ are chaos map
outputs for initial values $x_0$ and $y_0$.
Also,  $\varpi_{\tau_1,\tau_2}(x_0,y_0,r,n)$ denotes  $\varpi(x_0,y_0,r,n),$ which the
decimal part of the numbers of the first row is cut from the $\tau_1$-th decimal number to
the next digits and the decimal part of the numbers of the second row is cut
from the $\tau_2$-th decimal number to the next digits.
\end{rem}

\begin{rem}
The outputs of the decryption by reversible cellular automata for $\textbf{x}$
and $\textbf{y}$ as inputs in the $t-1$ and $t$ times, respectively, rule number
"$rule$" and after "$rep$" repetitions will be shown by
$[\textbf{x}^\prime,\textbf{y}^\prime]:=\Phi( \textbf{x},  \textbf{y}, rule, rep).$ Also,
the output of the decryption by irreversible cellular automata
for $\textbf{x}$ as inputs by  rule number "$rule$" and after "$rep$"
repetitions will be shown by $\textbf{x}^\prime:=\overline{\Phi}( \textbf{x}, rule, rep)$.
\end{rem}

\subsection{Key generation function}
One of the most important parts of an encryption algorithm is the key space.
In this subsection, we describe the method of choosing the keys used in the algorithm.
The keys used in the proposed algorithm are divided into three categories.
A number of keys are selected as inputs with a fixed value,
some of the keys are dependent on the input image and a number of
 keys are changed for each run of the algorithm.
Therefore, the key space for the different images will be different.
 Also, for a fixed image, the key space for each implementation of
 the algorithm will be different.
The following key space is used in the proposed algorithm.
\begin{align}
key=\{r_1,r_2,x^1_0,y^1_0,x^2_0,y^2_0\},
\end{align}
where parameters $r_1,r_2$ are consider as the parameter $r$, and $x^i_0,y^i_0~(i=1,2)$
 are considered as starting value in the chaotic systems.
 In the encryption algorithm, $r_1$ and $r_2$ are considered as fixed key, and
 the rest of the key space are
 generated by using key generation function.
 The following algorithm is used as key generation function (It should be noted that
matlab notation formats are used in the algorithms).\\
\begin{enumerate}
\item[\textbf{Step 1.}] Consider a color image  $I$ of the size $n \times m \times 3$,
a text `` \verb"Text" '' and inputs  $r_1,r_2$.
If $n$ is an odd number, then the last row of the plain image is repeated twice. In the next steps,
the number of rows (i.e., $n$) is considered as an even number.

\item[\textbf{Step 2.}] Matrix $\Psi$ of the size $n \times m \times 3$
based on the chaos map and image $I$ are obtained as follows:\\
\line(1,0){250}
\begin{algorithmic}
\For{$k=1:3$}
\For{$i=1:n$}
\State $\Psi(i,1,k)=\sum I(i,:,1)/(256\times m)$;
\EndFor
\For{$j=1:n/2$}
\State $\Psi(2j-1,2:m,k)=\varpi \big (\Psi(2j-1,1,k), \Psi(2j,1,k), r_1,m\big)(1,:);$
\State $\Psi(2j,2:m,k)=\varpi \big(\Psi(2j-1,1,k), \Psi(2j,1,k), r_1,m\big)(2,:);$
\EndFor
\EndFor
\end{algorithmic}
\line(1,0){250}\\
\item[\textbf{Step 3.}] By using the BITXOR operation, vector  $\xi_1$ and $\xi_2$
are found by using the following algorithm.\\
\line(1,0){250}
\begin{algorithmic}
\For{$k=1:3$}
\State $\Upsilon(:,:,k)= I(:,:,k)\oplus \Psi(:,:,k)$;
\For{$j=1:n-1$}
\State $\Upsilon(j+1,:,k)=\Upsilon(j,:,k)\oplus \Upsilon(j+1,:,k)$;
\EndFor
\For{$j=1:m-1$}
\State $\overline{\Upsilon}(:,j+1,k)=\overline{\Upsilon}(:,j,k)\oplus \overline{\Upsilon}(:,j+1,k)$;
\EndFor
\EndFor
\State $\xi_1=\bigoplus^3_{i=1}\Upsilon(n,:,:)$;
\State $\xi_2=\bigoplus^{3}_{i=1}\overline{\Upsilon}(:,m,:)$;
\end{algorithmic}
\line(1,0){250}\\

\item[\textbf{Step 4.}] The final vectors $v_1$ and $v_2$ are calculated as follows:
\begin{align*}
v_1=w(1,:),~~v_2=w(2,:),
\end{align*}
where
\begin{align*}
w=\varpi\Big(\sum \xi_1/(n\times 256\times 3),\sum \xi_2/(m\times 256\times 3), r_2,n\Big).
\end{align*}

\item[\textbf{Step 5.}] Key $x^1_0$ is considered as $\big(v_1+v_2\big)(1,n/2)/2$.
\item[\textbf{Step 6.}] By using $x^1_0$ and input text, the vector $T$ is
made by the following conditions.\\
First, $y_0$ is defined as

\begin{equation*}
y_0=mod\Big(\bigoplus_{i=1}^{text~size}\verb!Text!,1\Big),
\end{equation*}

$-$ If the input text size is $m$, $w$ is defined as
\begin{align*}
w(1,i):=
\left\{%
\begin{array}{ll}
\varpi\big(x^1_0,y_0,r_1,[m/2]+1\big)(1,[i/2]),&\textnormal{if $i$ is odd},\\
\\
\varpi\big(x^1_0,y_0,r_1,[m/2]+1\big)(2,i/2),&\textnormal{if $i$ is even},\\
\end{array}%
\right.
\end{align*}
then, we define
\begin{equation*}
T=\verb!Text!\oplus w(1,1:m).
\end{equation*}

$-$ If the input text size is smaller than $m$, then remaining parts
will be filled using the following vector elements:
\begin{align*}
w(1,i):=
\left\{%
\begin{array}{ll}
\varpi\big(x^1_0,y_0,r_1,[text~size-m/2]+1\big)(1,[i/2]),&\textnormal{if $i$ is odd},\\
\\
\varpi\big(x^1_0,y_0,r_1,[text~size-m/2]+1\big)(2,i/2),&\textnormal{if $i$ is even}.\\
\end{array}%
\right.
\end{align*}
$-$ If the input text size is bigger than $m$, then $T$ is found by

\begin{equation*}
T=\verb!Text!(1,1:m)\oplus \varpi(x^1_0,y_0,r_1,m)(1,:)\oplus \varpi(x^1_0,y_0,r_1,m)(2,:).
\end{equation*}

\item[\textbf{Step 7.}] The binary vector $T$ has been divided into $8$
section and $s_i,\delta_i~(i=1,\ldots,8)$ are found as:\\
\line(1,0){250}
\begin{algorithmic}
\For{$i=1:8$}
\State $s_i=circshift\Big(T\big(1,9(i-1):8i\big),[1,~5]\Big)$;
\State $\delta_i=\bigoplus^{m}_{j}s_i(1,j),~(i=1,\ldots,8)$;
\EndFor
\end{algorithmic}
\line(1,0){250}\\

\item[\textbf{Step 8.}] The vector $v_3$ is found  by using the following algorithm. \\
\line(1,0){250}
\begin{algorithmic}
\For{$i=1:8$}
\State $rule=mod \big (\varpi_{10,10}(\delta_i,\delta_{i+1},r_2,1),m\big)(1,1)$;
\State $rep=mod \big(\varpi_{10,10}(\delta_i,\delta_{i+1},r_2,1),m\big)(1,1)$;
\State $s^{\prime}_i=\overline{\Phi}(s_i,rule,rep)$;
\EndFor
\State $v_3=(s^{\prime}_1,s^{\prime}_2,s^{\prime}_3,s^{\prime}_4,s^{\prime}_5,
s^{\prime}_6,s^{\prime}_7,s^{\prime}_8)./256 $;
\end{algorithmic}
\line(1,0){250}\\

In the above algorithm, we consider $\delta_9=\delta_1$.

\item[\textbf{Step 9.}] $y^2_0$ is defined as
\begin{align*}
y^2_0=mod\Big(\sum^3_{i=1}v_i,1\Big).
\end{align*}
Also $x^2_0$ and $y^1_0$ are obtained as follows
\begin{align*}
&x^2_0=mod\big(x^1_0+\sum_i \varpi(x_r,y_r,r_2,m)(1,i),1\big),  \\
&y^1_0=mod\big(y^2_0+\sum_{i}\varpi(x_r,y_r,r_2,m)(2,i),1\big),
\end{align*}
where $x_r$ and $y_r$ denote random numbers in the interval $(0,1)$.
\begin{figure}
\centering
\hspace{-1.98cm}
\includegraphics*[height=6cm, width=12cm]{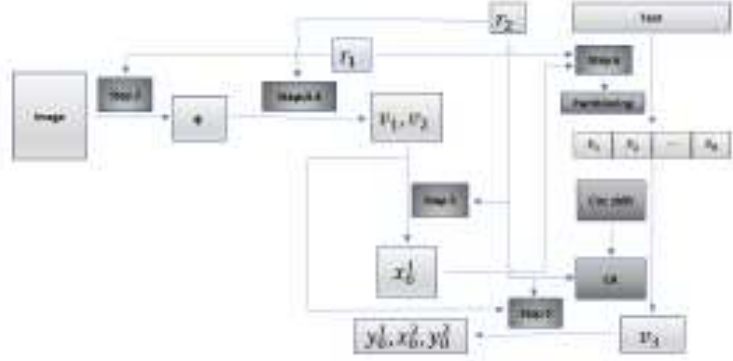}
\vspace{-1.0cm}
\emph{\caption{
 Proposed key generation algorithm.
}\label{f7}}
\end{figure}
\end{enumerate}
The flowchart of the proposed key function is given in Fig. \ref{f7}.
The above algorithm can also be performed for grayscale and binary images
by making some changes.
For grayscale and binary images, in steps 2-3, $k$ is considered equal to
$1$. Also, in step 4, $w$ is changed as
\begin{align*}
w=\varpi\Big(\sum \xi_1/(n\times 256),\sum \xi_2/(m\times 256), r_2,n\Big).
\end{align*}
To check the key space sensitivity, numerical results for the above
algorithm are presented in Table \ref{t1}.
\begin{center}
\begin{table}[ht!]
\footnotesize
\caption{ {\footnotesize Key space generated by the proposed algorithm for various images and texts.}}
\label{t1}
\centering
\begin{tabular}{cccccccccccc}
\hline
          \multicolumn{3}{c}{}   &\multicolumn{6}{c}{Keys }   \\\cmidrule{5-10}
Image& \thead{Number of \\ key space}  & &Text     &$r_1$     &$r_2$                  &$x^1_0$     &$y^1_0$   &$x^2_0$&$y^2_0$     \\
\hline \hline
girl     &1&  Original        &\verb"Hello, world"   &1.2000     &0.7500     &0.5279  &0.5512    &0.4427   &0.2714     \\
$(256\times256\times3)$  &2&  One bit changed &\verb"Hello, world"                    &1.2000     &0.7500     &0.8815  &0.1107    &0.2561   &0.7083     \\
           & 3      &  Original        &\verb"Hello world"                    &1.2000     &0.7500     &0.5279  &0.5127    &0.7928   &0.8495      \\
           &4&  Original &\verb"Hello world"                            &1.0000     &0.5000     &0.6080  &0.0216    &0.4172   &0.6206     \\
           \hline
lena    &5  &  First time run        &\verb"New world"                  &1.0000     &0.5000    &0.6019  &0.4763    &0.6122   &0.9951       \\
$(256\times256\times3)$   &6&  Second time run   &\verb"New world"      &1.0000     &0.5000    &0.6019  &0.4488    &0.8778   &0.9951    \\
           &7&Third time run          &\verb"New world"                 &1.0000     &0.5000    &0.6019  &0.0902    &0.5927   &0.9951      \\
&8&  One bit changed &\verb"New world"                                  &1.0000     &0.5000    &0.1974  &0.5202    &0.1163   &0.6208     \\
&9&  Original &\verb"new world"                                         &1.0000     &0.5000    &0.6019  &0.8281    &0.4053   &0.4990     \\
&10&  Original &\verb"newworld"                                         &1.0000     &0.5000    &0.6019  &0.7584    &0.6434   &0.3506     \\
           \hline
pirate         &11 &  First time run       &\verb"Hello world"           &0.9000     &0.7000    &0.5799  &0.0129    &0.0525  &0.5215       \\
$(256\times256)$           &12&   Second time run &\verb"hello world"    &0.9000     &0.7000    &0.5799  &0.4723    &0.9962  &0.5215      \\
           &13& Third time run         &\verb"Hello world"               &0.9000     &0.7000    &0.5799  &0.1153    &0.8749  &0.5215       \\
           &14&  One bit changed &\verb"Hello world"                     &0.9000     &0.7000    &0.7090  &0.2551    &0.9203  &0.4178     \\
          &15&  Original &\verb"HellO world"                             &0.9000     &0.7000    &0.5799  &0.6393    &0.8851  &0.9122     \\
           \hline
deer       &16&  Original        &\verb"Newworld "                       &0.5000     &1.2000    &0.1543    &0.9777   &0.2099  &0.6800        \\
$(256\times256)$&17&  Original        &\verb"New worlD "                      &0.5000     &1.2000    &0.1543    &0.0793   &0.4856  &0.3792        \\
           &18&  Original        &\verb"New world"                       &0.5000     &1.2000    &0.1543    &0.2571   &0.9620  &0.3675    \\
           &19&  One bit changed &\verb"New world"                       &0.5000     &1.2000    &0.6818    &0.5516   &0.8778  &0.9818   \\
           &20&  One bit changed &\verb"Newworld"                        &0.5000     &1.2000    &0.6818    &0.5968   &0.3163  &0.2943   \\
           \hline
parthenon   &21& First time run        &\verb"Hello, world"                  &0.7500     &1.2000    &0.6873    &0.2154   &0.0757  &0.3833        \\
$(256\times256)$&22&  Second time run        &\verb"Hello, world"                  &0.7500     &1.2000    &0.6873    &0.6651   &0.0649 &0.3833        \\
\hline
\end{tabular}
\end{table}
\end{center}
\subsection{Encryption process}
In this section, a color image of the size $n\times m\times 3$ is considered as
an input image and the proposed algorithm for image encryption based on the 2D-hybrid
chaos map and cellular automata is written as follows:
\begin{enumerate}
\item[ \textbf{Step1.}] Input image is divided into its own three plates.
\item[\textbf{Step2.}] First the odd columns of each plates are moved to the
left hand and so the even columns are moved to the right hand, then the odd
rows of each plates are moved up and so the even rows are moved down. By
applying this processes twice, each plates are divide into sixteen segments
and each parts can see as the small size of the input image.
The output images of this step is seen as blocks matrix of the size
$4\times 4.$  We show these block matrices by $I_{\mathsf{b}}^k,\: k=r,g,b.$
\item[\textbf{Step3.}] The mixed block matrix $I_{\mathsf{b}}^{\mathfrak{m}}$ of
the size $6\times 8,$ is achieved by using the following algorithm:\\

\line(1,0){250}
\begin{algorithmic}{
\For{$j=1:2$}{
\State $I_{\mathsf{b}}^{\mathfrak{m}}(:,2j-1)=I_{\mathsf{b}}^r(:,j);$
\State $I_{\mathsf{b}}^{\mathfrak{m}}(:,2j)=I_{\mathsf{b}}^b(:,j);$
\State $I_{\mathsf{b}}^{\mathfrak{m}}(j+4,:)=\big(I_{\mathsf{b}}^g(2j-1,:),I_{\mathsf{b}}^g(2j,:)\big);$
\EndFor}}
\end{algorithmic}
\line(1,0){250}\\

where $I_{\mathsf{b}}^k(i,j), (k=r,g,b,\mathfrak{m})$ denotes $(i,j)$-th block of
the block matrix $I_{\mathsf{b}}^k.$
\item[\textbf{Step4.}] First the odd and even columns of block
matrix $I_{\mathsf{b}}^{\mathfrak{m}},$ are shifted $\varpi_{1,1}(x_0,y_0,r,8)(1,j)$
units up and down, respectively, then the odd and even rows of
$I_{\mathsf{b}}^{\mathfrak{m}},$ are shifted $\varpi_{1,1}(x_0,y_0,r,8)(2,j)$
units right and left, respectively. We show this obtained shifted
mixed matrix by $SI_{\mathsf{b}}^{\mathfrak{m}}.$
\item[\textbf{Step5.}] By using the following algorithm, the block
matrix $SI_{\mathsf{b}}^{\mathfrak{m}}$ is separated into three matrices, named $L,C$ and $R:$\\

\line(1,0){250}
\begin{algorithmic}
\State $L=SI_{\mathsf{b}}^{\mathfrak{m}}(1:4,1:4);$
\State $R=SI_{\mathsf{b}}^{\mathfrak{m}}(1:4,5:8);$
\State $C=\big(SI_{\mathsf{b}}^{\mathfrak{m}}(5,1:4);
 SI_{\mathsf{b}}^{\mathfrak{m}}(5,5:8);SI_{\mathsf{b}}^{\mathfrak{m}}(6,1:4);
 SI_{\mathsf{b}}^{\mathfrak{m}}(6,5:8)\big);$
\end{algorithmic}
\line(1,0){250}\\

\item[ \textbf{Step6.}] We take the matrices $L,C,R$, all together and carry out the
shifting operation of the arrays of these matrices as follows:\\
We shift the matrix $L$  in a circular fashion (clockwise) and starting from the
array $L(1,1).$ Note that in a shifting of the arrays of the matrix $L,$ if we
reach the last array in the shifting process, like the array $L(e_r,e_c),$
we will enter in the matrix $C$ from the array $C(e_r,e_c),$ and continue shifting
in a circular fashion (counter clockwise) and when we reach the array $C(1,1),$ we
will enter in the matrix $R$ from the array $R(1,1)$ and continue shifting in a
circular fashion (clockwise) and if we reach the $R(e_r,e_c),$ we will enter in
the matrix $L$ from the array $L(1,1)$ and continue shifting again clockwise
(see Fig. \ref{f8}). We show these three obtained matrices by $L_s, C_s$ and $R_s$,
respectively.
\begin{figure}
\centering
\hspace{-1.98cm}
\includegraphics*[width=.70\textwidth]{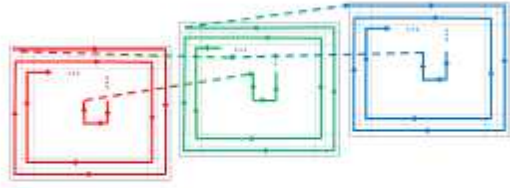}
\vspace{-2cm}
\emph{\caption{
 Proposed circular fashion shifting process.
}\label{f8}}
\end{figure}
\item[\textbf{Step7.}]
First of all, if $n$ is odd number, we repeat the last row of each matrices
$L_s, C_s$ and $R_s,$ one more time in the last of its, and so the number of
the row will be even. Now, we continue our encryption algorithm by
applying the cellular automata as follow (see Fig. \ref{F9}):
\begin{enumerate}
\item[(i)] All of the matrices $L_s, C_s$ and $R_s$ are converted to the binary matrices.
\item[(ii)] The rows of each matrices are segmented to 8 parts and are
considered as inputs of $\Phi$
in the following way. For $K=L,C,R,$ we have\\
\begin{figure}\label{Circ}
\centering
\hspace{-1.98cm}
\includegraphics*[width=.70\textwidth]{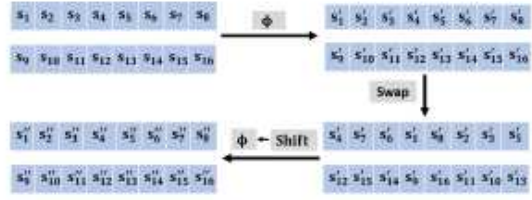}
\vspace{-0.7cm}
\emph{\caption{
Process of Step 7.
}\label{F9}}
\end{figure}
\\

\line(1,0){250}
\begin{algorithmic}
\For{$i=1:n/2$}
\For{$k=1:8$}
\State $\textbf{x}=K_s\big(i,(k-1)m+1:km\big);$
\State $\textbf{y}=K_s\big(n/2+i,(k-1)m+1:km\big);$
\State $rule=\varpi_{1,2}(x_0, y_0, r, n/2)(1,i);$
\State $rep=\varpi_{1,2}(x_0, y_0, r, n/2)(2,i);$
\State $[\textbf{x},\textbf{y}]=\Phi( \textbf{x},  \textbf{y}, rule, rep);$
\EndFor
\EndFor
\end{algorithmic}
\line(1,0){250}\\

\item[(iii)] The rows of each matrices was segmented to
8 parts and swaped in the following way:

\line(1,0){250}
\begin{algorithmic}
\For{$i=1:n$}
\For{$k=1:8$}
\State $seg_k=K_s\big(i,(k-1)m+1:km\big);$
\EndFor
\State $[seg_1, seg_4]=swap(seg_1,seg_4);$
\State $[seg_2,seg_7]=swap(seg_2,seg_7);$
\State $[seg_3,seg_6]=swap(seg_3,seg_6);$
\State $[seg_5,seg_8]=swap(seg_5,seg_8);$
\EndFor
\end{algorithmic}
\line(1,0){250}\\

\item[(iv)] By using the function "$circshift$",  the array of the matrices is
shifted to the right hand in the following way:\\

\line(1,0){250}
\begin{algorithmic}
\For{$i=1:n/2$}
\State $v=[K_s(i,:), K_s(n/2+i,:)];$
\State $t=\varpi_{4,0}(x_0, y_0, r, n/2)(1,i);$
\State $v=circshift(v,t);$
\State $K_s(i,:)=v(1,1:m);$
\State $K_s(n/2+i,:)=v(1,m+1:2m);$
\EndFor
\end{algorithmic}
\line(1,0){250}\\

\item[(v)] The Step(7)-(a) is repeated one more time.
\end{enumerate}
We show these three obtained matrices by $L^r_1,L^g_1$ and $L^b_1,$ respectively.
\item[ \textbf{Step8.}] In this step, $L^r_2,L^g_2$ and $L^b_2$ are calculated by
using the following algorithm\\
\line(1,0){250}
\begin{algorithmic}
\State $x_0=(x^1_0+x^2_0)/2;$ $y_0=(y^1_0+y^2_0)/2;$ $r=\max(r_1,r_2);$
\State $Q^r=\varpi\big(x_0,y_0,r,n\times m\big)(1,:);$
\State $Q^g=\varpi\big(x_0,y_0,r,n\times m\big)(2,:);$
\State $Q^b=Q^r\oplus Q^g;$
\For{$j=r,g,b$}
\State $L^j_2=Q^j\oplus L^j_1;$
\EndFor
\end{algorithmic}
\line(1,0){250}\\

\item[ \textbf{Step9.}] At the last step, we put $L^r_2,L^g_2$ and $L^b_2$
on each other as plates of the final encrypted image.
\end{enumerate}
\begin{rem}
To encrypt a grayscale or a binary images, we can apply the steps 2, 6, 7 and 8.
Just in the step 6,
 if we reach the last array in the shifting process, like the array $I(e_r,e_c),$
we will come back to the first array, i.e. $I(1,1),$ and continue shifting in a
circular fashion (clockwise).
As an example for step 6,
in Fig. \ref{F10}\textcolor{blue}{.b-c}, an image where almost the center of it is changed
 to white, is used to simulate, and Fig. \ref{F10}\textcolor{blue}{.a}
 shows circular fashion shifting process for grayscale image.
\end{rem}

\begin{figure}
\centering
\hspace{-1.98cm}
\includegraphics*[width=.90\textwidth]{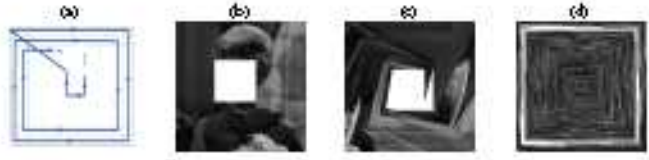}
\vspace{-1.5cm}
\emph{\caption{
Showing the results of the mixing of an image by circular fashion shifting: (a) proposed shift,
(b) given the image, (c) circular fashion shifting after $190$ shifts
 (d) circular fashion shifting after $150000$ shifts.}\label{F10}}
\end{figure}
\section{{\textbf Steganography}}
\begin{figure}
\centering
\subfigure{
\includegraphics*[width=.70\textwidth]{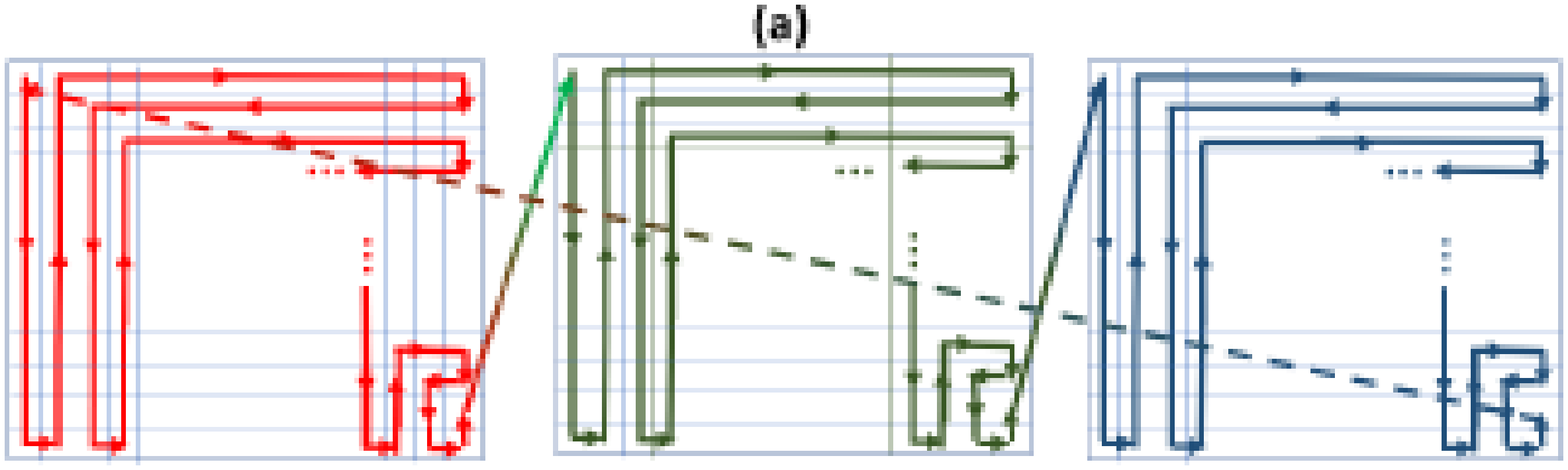}}\\
\vspace{-0.90cm}
\subfigure{
\includegraphics*[width=.70\textwidth]{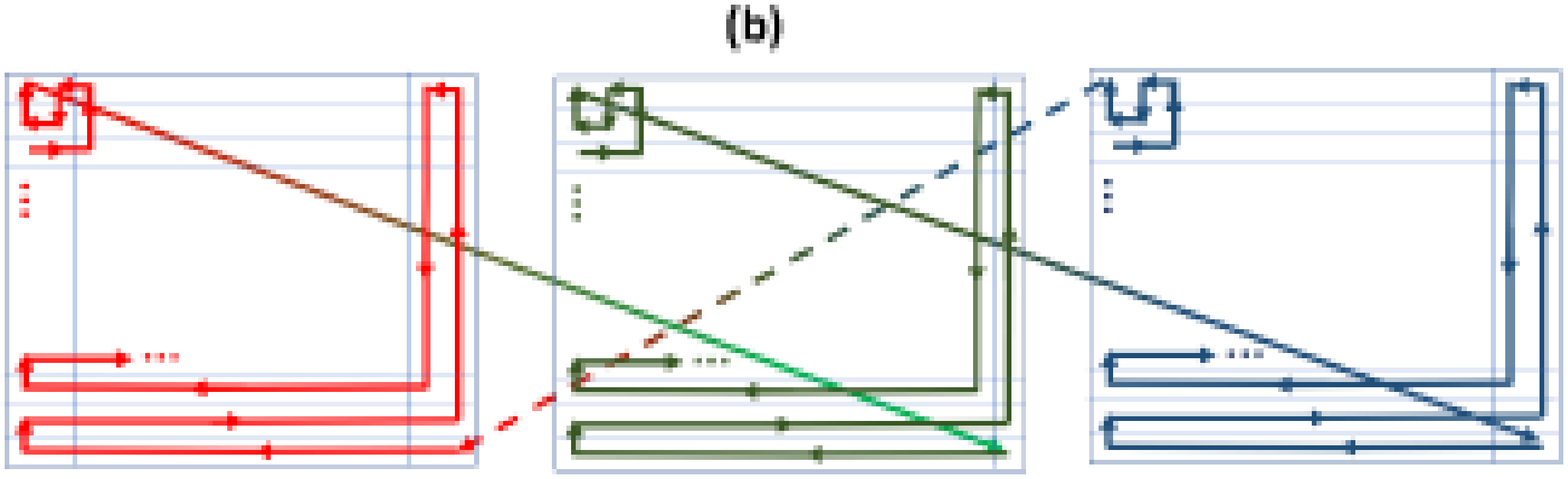}}\\
\vspace{-1.0cm}
\emph{\caption{
Proposed spiral fashion shifting process for a color image:(a) shift I, (b) shift II.
}\label{F11}}
\end{figure}

In this section, $I$ is a RGB image of the size $n\times m\times 3$ and $w$
is considered as
a secret image of the size $n^{\prime}\times m^{\prime}$,
where $n^{\prime}m^{\prime}\leq nm/3$.
\begin{enumerate}
\item[\textbf {Step1.}] Input a cover image $I$ and a secret image $w.$
\item[ \textbf{Step2.}] Image $I$ is divided into its own three plates and
named them $I^k, k=r,g,b.$
\item[\textbf {Step3.}] To avoid exorbitant shifts in the arrays of
the matrix $I,$ we use the function $mod$.
We take the matrices $I^k, k=r,g,b$ all together and carry out the
shifting operation of the arrays of these matrices as follows:\\
First, we shift the matrix $I^r$  in a spiral fashion and starting
from the array $I^r(1,1)$ and continuing along the column of the matrix
and the row of the matrix, respectively. Note that in a shifting of the
arrays of the matrix $I^r,$ if we reach the last array of this matrix in
the shifting process,
we will enter in the matrix $I^g$ from the array $I^g(1,1)$ and continue
shifting in the same way and when we reach the last array of this matrix
in the shifting process, we will enter in the matrix $I^b$ from the array
$I^b(1,1)$ and continue shifting in a the same way and if we reach the last
array of this matrix, we will come back to the matrix $I^r$ from the array
$I^r(1,1)$ and continue shifting. We called this shift $I.$ (See
Fig. \ref{F11}\textcolor{blue}{.(a)}).
Then, we considered these three obtained matrices again all together and start
shifting similar to the above way, but starting from the last array $I^r(n,m)$
and continuing along the row of the matrix and the column of the matrix,
respectively. We called this shift $II$ (see Fig. \ref{F11}\textcolor{blue}{.(b)}).
We denote these three obtained matrices by $I^k_s, k=r,g,b.$
\item[\textbf {Step4.}]
\begin{enumerate}
\item[(i)] We get the matrices $I^k_{s,d}, k=r,g,b$ by applying
the discrete framelet transform on the matrices $I^k_{s}, k=r,g,b$, and we cut
the $LL$ part of $I^k_{s,d}, k=r,g,b$, and named them $\widehat{I}^{k}_{s,d},  k=r,g,b.$
\item[(ii)] In this step $n^{\prime\prime}$ and $m^{\prime\prime}$ are considered as row and column
numbers of the output matrices in the part (i).
 First the odd and even columns of
matrices $\widehat{I}^{k}_{s,d},  k=r,g,b$, are shifted $\varpi_{3,3}\big(x_0,y_0,r,\max(n^\prime\prime,m^\prime\prime)\big)(1,j)$
units up and down, respectively, then the odd and even rows of
$I_{\mathsf{b}}^{\mathfrak{m}}$ are shifted $\varpi_{3,3}\big(x_0,y_0,r,\max(n^\prime\prime,m^\prime\prime)\big)(2,j)$
units right and left, respectively.
\end{enumerate}

\item[\textbf {Step5.}]
By considering the arrays of the matrices $\widehat{I}^{k}_{s,d}, k=r,g,b$
and $w$ as interger numbers, we converted these matrices to the binary
matrices named $b\widehat{I}^{k}_{s,d}, k=r,g,b$ and $bw,$ respectively.
We set the most significant bits of the $bw$ in the least significant
bits of the $b\widehat{I}^{k}_{s,d}, k=r,g,b$ as the following algorithm
and the obtained matrices $b\widehat{I}^{k}_{s,d}, k=r,g,b$, is converted
to the matrices with integer arrays.\\
\line(1,0){250}
\begin{algorithmic}
\For{$i=1:n^\prime$}
\For{$j=1:m^\prime$}
\State $b\widehat{I}^{r}_{s,d}(i,8j)=bw\big(i,8*(j-1)+1\big);$
\State $b\widehat{I}^{g}_{s,d}(i,8j)=bw\big(i,8*(j-1)+2\big);$
\State $b\widehat{I}^{g}_{s,d}(i,8j-1)=bw\big(i,8*(j-1)+3\big);$
\State $b\widehat{I}^{b}_{s,d}(i,8j)=bw\big(i,8*(j-1)+4\big);$
\EndFor
\EndFor
\end{algorithmic}
\line(1,0){250}\\
\item[\textbf {Step6.}] First, the $LL$ of the ${I}^{k}_{s,d}, k=r,g,b$, is replaced by
$b\widehat{I}^{k}_{s,d}, k=r,g,b$ and then, the inverse discrete framelet transform is
applied on the achieved matrices and we put these matrices as the plates of the final image.
\end{enumerate}

\begin{figure}
\centering
\includegraphics*[width=.70\textwidth]{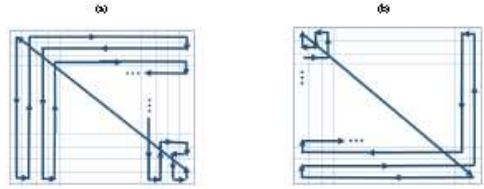}
\vspace{-0.5cm}
\emph{\caption{
Proposed spiral fashion shifting process for a grayscale image.
}\label{F12}}
\end{figure}

Similarly, we can do all the steps relevant to the forward part in the
opposite way to get the secret image.
\begin{rem}
For the steganography process of a grayscale or a binary images in a
grayscale images, we can apply the steps 1, 3, 4, 5 and 6.
Just in the steps 3 and 6,
 if we reach the last array in the shifting process,
we will come back to the first array, i.e. $I(1,1),$ and continue shifting in a spiral fashion (see Fig. \ref{F12}).
(For more examples, we refer to Appendix section).
Also, step 5 for grayscale in grayscale and binary in grayscale are as follows:
\\

$-$Grayscale in grayscale:\\
\line(1,0){250}
\begin{algorithmic}
\For{$i=1:n^\prime$}
\For{$j=1:m^\prime$}
\For{$k=1:4$}
\State $b\widehat{I}_{s,d}(i,8j-4+k)=bw(i,8*(j-1)+k);$
\EndFor
\EndFor
\EndFor
\end{algorithmic}
\line(1,0){250}\\

$-$Binary in grayscale:\\
\line(1,0){250}
\begin{algorithmic}
\For{$i=1:n^\prime$}
\For{$j=1:m^\prime$}
\State $b\widehat{I}_{s,d}(i,8j)=w(i,j);$
\EndFor
\EndFor
\end{algorithmic}
\line(1,0){250}\\
\end{rem}
\section{Numerical experiments and security analysis}
In this section, simulation results for the proposed algorithms are given.
In the numerical results for space key, we use Table \ref{t1}. Note that, case (iii)
is used as the chaos map in the proposed algorithm.
Also, in order to show the performance of the proposed algorithms, the results
are compared with the results of other papers.
\subsection{Simulation results of the encryption algorithm}
\begin{figure}
\centering
\includegraphics*[width=.80\textwidth]{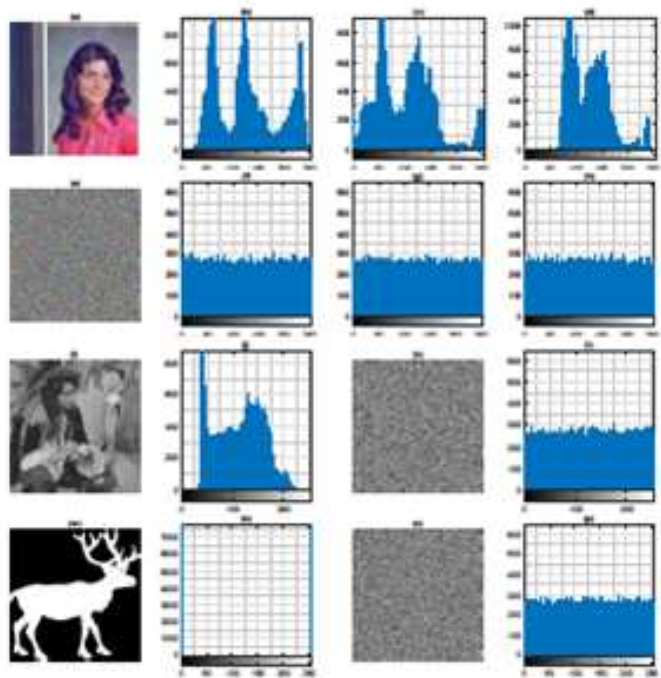}
\emph{\caption{
Original images and encryption results and their histograms: (a)-(d) girl original image,
(e)-(h) encrypted girl image, (i)-(j) pirate original image, (k)-(l) encrypted pirate image,
(m)-(n) deer original image, (o)-(p) encrypted deer image.
}\label{F77}}
\end{figure}

The security analysis of the proposed algorithm are studied in this subsection.
Also, the results for encryption proposed algorithm are given in Fig. \ref{F77}.

 \subsubsection{Security key space}
\begin{figure}
\centering
\subfigure{
\includegraphics*[width=.90\textwidth]{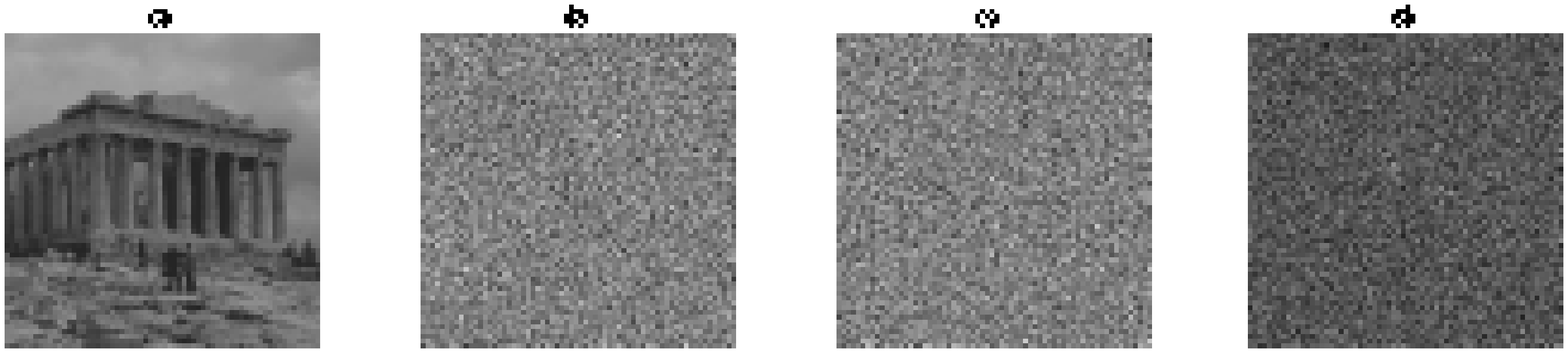}}\\
\vspace{-1.00cm}
\subfigure{
\includegraphics*[width=.90\textwidth]{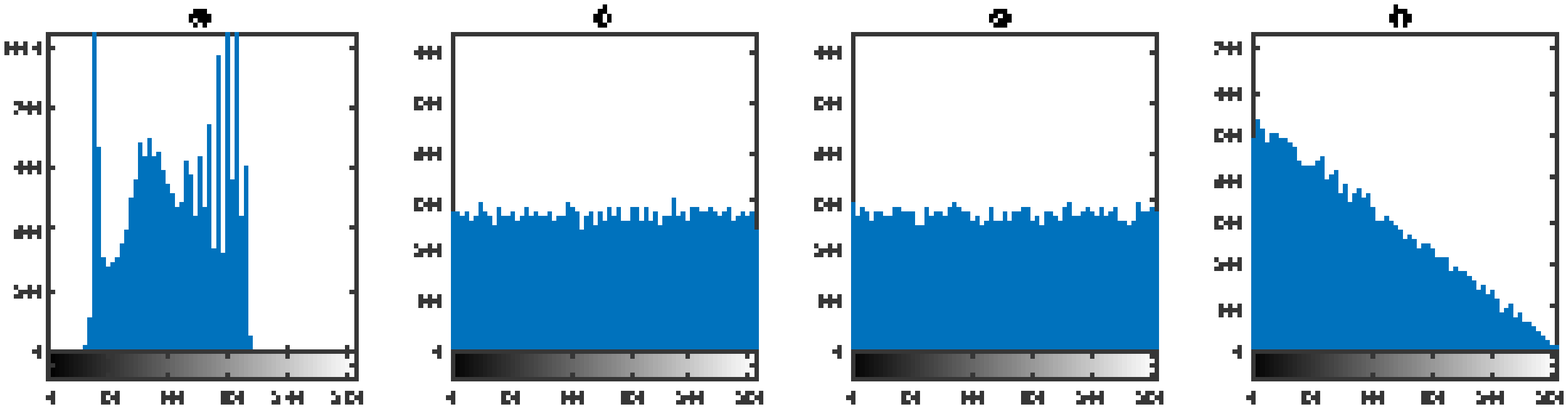}}\\
\vspace{-0.25cm}
\emph{\caption{ Results for image and histogram:
(a-e) the plain image, (b-f) the first encrypted image,
(c-g) the second encrypted image, (d-h) the pixel-to-pixel image.
}\label{F14}}
\end{figure}

\begin{figure}
\centering
\includegraphics*[width=.90\textwidth]{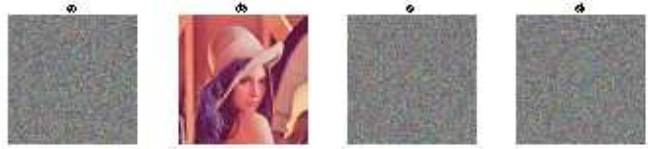}
\vspace{-0.5cm}
\emph{\caption{
Decryption process for different keys.}\label{F15}}
\end{figure}

 One of the most important components of the encryption algorithm is key space.
 The key space in the proposed algorithm has a dynamic structure i.e., when a bit of
 the input image is changed, the key space changes,
 also, a different key space is used for each algorithm implementation.
 In Fig. \ref{F14}, the proposed algorithm is run twice on the parthenon image.
In these results, the key spaces 21 and 22 from Table \ref{t1} are used.
To show the difference between the two images, pixel-to-pixel difference is given in Fig. \ref{F14}.
The results show, the two images are very different.
Therefore, it can be concluded that the proposed algorithm is able to
resist the chosen-plain text attack.
Also, key size is important item in key space.
Key size plays an important role in attacks.
The relationship between key size and brute force
attack has been studied in \cite{42}.
According to this study, if key space is larger than $2^{100}$, then the algorithm
can withstand the brute force attack.
The key space in the proposed algorithm consists of six keys.
Therefore, if in the
image encryption algorithm,
the precision of the key size is equal to $10^{-15}$, then the key space is
$10^{90}=2^{90\log_210}\approx 2^{298}$, therefore
the key space for resistance against the brute force attack will be sufficient enough.
As last study in this subsection,
the key space changes is studied in the decryption process.
By applying small changes on the key space 5 from Table \ref{t1},
two keys are created as follows
\begin{align*}
&\overline{K}_5=\{1,0.5,0.6019+10^{-15},0.4763,0.6122,0.9951\},\\
&\widehat{K}_5=\{1,0.5,0.6019,0.4763,0.6122+10^{-15},0.9951\},
\end{align*}
In Fig. \ref{F15}\textcolor{blue}{.a}, lena image is encrypted by $K_5$
(key space number 5, from Table \ref{t1}), in Fig. \ref{F15}\textcolor{blue}{.b},
the decryption process is run by $K_5$, also in the Fig.s\ref{F15}\textcolor{blue}{.c-d},
the decryption process are run
by $\overline{K}_5$ and $\widehat{K}_5$, respectively.
Therefore by small changes in the keys, we can not decrypt the image.
Table \ref{table:t2} shows that the key space size of the proposed algorithm
is bigger than the key space size of the compared algorithms.
\begin{center}
\begin{table}[ht!]
\caption{ {\footnotesize Comparison of the key space size.}}
\label{table:t2}
\centering
\begin{tabular}{ccccccccc}
\hline
&Proposed algorithm  &Ref.\cite{37}   &Ref.\cite{38}  &Ref.\cite{39}         \\
\hline\hline
Key space size&$10^{90}\approx 2^{298}$  &$10^{84}$  &$2^{120}$  &$10^{56}$         \\
\hline
\end{tabular}
\end{table}
\end{center}

\subsubsection{Statistical analysis}
\begin{figure}
\centering
\includegraphics*[width=.80\textwidth]{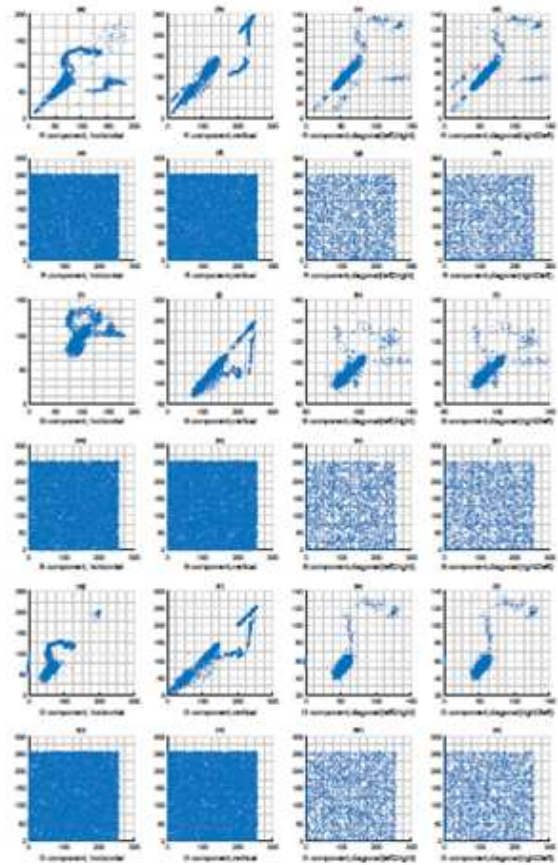}
\emph{\caption{
Correlation of neighbourhood pixels at different directions
before and after encryption of girl image.
}\label{F16}}
\end{figure}

In this subsection, statistical analysis including correlation values,
information entropy and histogram analysis are studied.
The correlation values for image can be found by using following formula from \cite{47}
\begin{align}
C_{x,y}=\frac{E(x-\mu_x)(y-\mu_y)}{\sigma_x\sigma_y},
\end{align}
where $E,\mu$ and $\sigma$ denote expectation, mean values and standard deviation,
respectively.
The correlation values for different images are given in Table \ref{table:t3}. By using
results in this table, we
can see that the original images have
high correlation values, while encrypted images have low correlation values.
Also, Fig. \ref{F16} shows correlation distributions for the original images and encrypted images.
Comparison results of the proposed algorithm with the algorithm in
\cite{40}, that are given in Table \ref{table:t4}, illustrates the advantage of the proposed
algorithm.
The information entropy is calculated by
\begin{align}
H(k)=-\sum^{w-1}_{i=0}P(k_i)\log_2P(k_i),
\end{align}
where $w$ and $P$ represent the gray level and the probability, respectively.
The values of this expression are in the interval $[0,8]$, and
the ideal value for the encrypted image is equal to 8.
Numerical results of the information entropy for different images are shown in Table \ref{table:t3}.
By using these results, we can say that the information entropy for
encrypted images are close to the ideal value. Also,
the comparison results for the information entropy are presented in
Table \ref{table:t5}.
Histogram diagram for original and encrypted images are given
in Fig. \ref{F77}. In this figure, we can see that encrypted images have a flat
distribution compared to the original images.
\begin{center}
\begin{table}[ht!]
\footnotesize
\caption{{\footnotesize Numerical results for UACI, NPCR, Information entropies,
Correlation coefficients for different images.}}
\label{table:t3}
\centering
\begin{tabular}{ccccccccc}
\hline
       \multicolumn{5}{c}{}   &\multicolumn{3}{c}{Correlation coefficients }   \\\cmidrule{6-9}
Image  &&UACI &NPCR&Information          &Horizontal      &Vertical &Diagonal        &Diagonal \\
&     && & entropies                 &                    & &\tiny{ (lower left to top right)}&\tiny{(lower right to top left)}\\
\hline \hline
                      &R&   &     &7.2549       &0.9786      &0.9879   &0.9688       &0.9684     \\
 girl-original        &G&  &     &7.2704       &0.9660      &0.9820    &0.9515      &0.9507    \\
                      &B&  &     &6.7825       &0.9523      &0.9718    &0.9307      &0.9306   \\
\hline
                      &R&33.5542   & 99.5850    &7.9971       &0.0028      &0.0088    &-0.0071       &0.0048     \\
girl-encrypted        &G&33.3360   & 99.5804    &7.9973       &-0.0014      &-0.0009    &-0.0006       &-0.0047    \\
                      &B&33.2780   & 99.5895    &7.9972       & -0.0004     &-0.0034    &-0.0086     &-0.0050   \\
\hline
pirate-original        & &   &  &7.2887       &0.9434      &0.9564    &0.9201       &0.9134   \\
\hline
pirate-encrypted        &&33.5110   & 99.5636    &7.9971       &-0.0014      &0.0006    &-0.0014       &-0.0036    \\
\hline
deer-original        & &   &   &0.9102       &0.9463      &0.9538    &0.9274       &0.9317   \\
\hline
deer-encrypted        &  &33.1739   & 99.6170  &7.9975       &0.0055      &-0.0009    &-0.0004       &-0.0068    \\
\hline
\end{tabular}
\end{table}
\end{center}

\begin{center}
\begin{table}[ht!]
\caption{ {\footnotesize Correlation coefficients for
lena ($256 \times 256 \times 3$)
by 10,000 pixels pairs of adjacent positions.}}
\label{table:t4}
\centering
\begin{tabular}{ccccccccc}
\hline
                      &     &Horizontal    &Vertical &Diagonal        \\
\hline\hline
Proposed algorithm     & R    &0.0038   &0.0094&0.002\\
                       & G    &-0.0062   &-0.0042&-0.0016\\
                       & B    &0.0109   &-0.0054 &-0.00695\\
                       \hline
Algorithm in \cite{40} & R    &-0.0127   &0.0067 &0.0060\\
                       & G    &-0.0075  &-0.0068&-0.0078\\
                       & B    &-0.0007  &0.0042 &0.0026\\
\hline
\end{tabular}
\end{table}
\end{center}

\begin{center}
\begin{table}[ht!]
\caption{ {\footnotesize Information entropy of the encrypted
color image for lena ($256 \times 256 \times 3$).}}
\label{table:t5}
\centering
\begin{tabular}{ccccccccc}
\hline
                 &&R   & G & B        \\
\hline\hline
Proposed algorithm &           &7.9970   &7.9973 &7.9973\\
Algorithm in \cite{32} &       &7.9891   &7.9898 &7.9899\\
Algorithm in \cite{33} &      &7.9896   &7.9893 &7.9896\\
Algorithm in \cite{34} &      &7.9893   &7.9896 &7.9903\\
Algorithm in \cite{35} &      &7.9874  &7.9872 &7.9866\\
Algorithm in \cite{36} &      &7.9278  &7.9744 &7.9705\\
\hline
\end{tabular}
\end{table}
\end{center}

\subsubsection{Sensitivity analysis}
NPCR and UACI are two important tools in sensitivity analysis.
NPCR (i.e., number of pixels change rate) represents
the number of pixels change rate while one pixel of plain image changed and UACI
(i.e., unified average changing intensity) denotes
the average intensity of difference between the plain image
and encrypted image.
A good study based on statistical discussions for these values is presented in \cite{43}.
The ideal value for NPCR and UACI are considered as $100\%$ and  $33.\overline{33}\%$,
respectively.
Using these values,  the resistance against  differential
and plaint text attacks can be studied.
When these values are close to the ideal values, we can say that
the encryption
algorithm is more sensitive to the changing of plain image,
therefore the encryption algorithm can effectively resist plain text
and differential attacks.
By using numerical results in Table \ref{table:t3},
we can see that results are close to the
ideal value.
As well as, the comparison results for NPCR and UACI
values have been tabulated in Table \ref{table:t6}.
In computing,  NPCR and UACI are considered as follows
\begin{align}\label{ex1}
&NPCR=\frac{\sum_{i,j}D(i,j)}{m\times n}\times 100\%,\\
&UACI=\frac{\sum_{i,J}|C_1(i,j)-C_2(i,j)|}{255\times m\times n}\times 100\%,
\end{align}
with
\begin{align}\label{ex1}
&D(i,j):=\left\{%
\begin{array}{ll}
1,&\textnormal{when}~C_1(i,j)\neq C_2(i,j),\\
\\
0,&\textnormal{when}~C_1(i,j)= C_2(i,j),\\
\end{array}%
\right.
\end{align}
where, $C_1$ and $C_2$ are considered as the encrypted
image before and after the original image
is changed.

\begin{center}
\begin{table}[ht!]
\caption{ {\footnotesize UACI and NPCR of the encrypted
color image for lena ($256 \times 256 \times 3$).}}
\label{table:t6}
\centering
\begin{tabular}{cccccccccccc}
\hline
&                   &R   & G & B        \\
\hline\hline
NPCR &Proposed algorithm           &99.59260   &99.63990 &99.60020\\
     &Algorithm in \cite{29}       &99.58649   &99.21722 &98.84796\\
     &Algorithm in \cite{30}       &99.42   &99.60 &99.54\\
     &Algorithm in \cite{31}       &99.26   &99.45 &99.13\\
     \hline
UACI &Proposed Algorithm             &33.49960   &33.41470   &33.41000\\
     &Algorithm in \cite{29}         &33.48347   &33.46399   &33.26891 \\
     &Algorithm in \cite{30}         &27.78      &27.66      &24.94\\
     &Algorithm in \cite{31}         &21.41      &23.42      &15.08\\
\hline
\end{tabular}
\end{table}
\end{center}

\subsubsection{Noise and data loss attacks}
\begin{figure}
\centering
\includegraphics*[width=.80\textwidth]{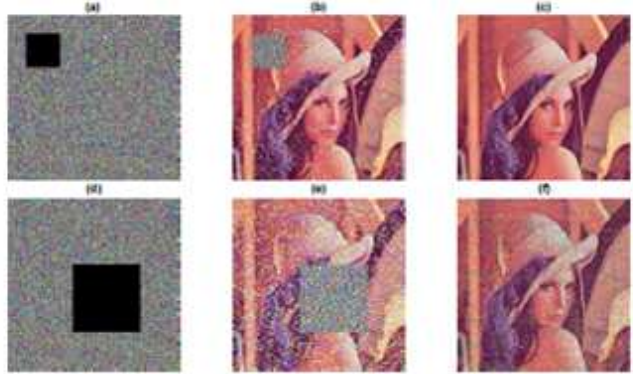}
\emph{\caption{
Results of the data loss attacks: (a),(d) attacked images,
(b),(e) results of the algorithm in \cite{41}, (c),(f) results of the
proposed algorithm.
}\label{F17}}
\end{figure}

\begin{figure}
\centering
\includegraphics*[width=.80\textwidth]{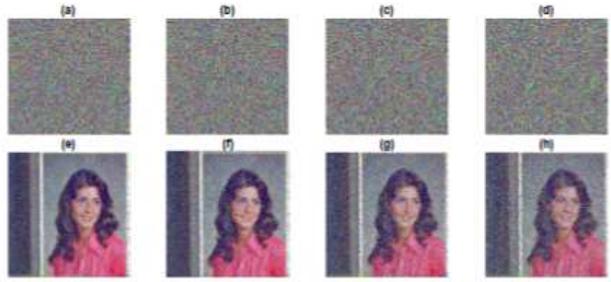}
\emph{\caption{Noise attack results with salt and pepper noise for
different noise density:
(a)-(d) noise attacks for encrypted image
with noise density $=0.08,0.1,0.15,0.2$, respectively. (e)-(h) decryption
results for noise attacks
with noise density $=0.08,0.1,0.15,0.2$, respectively.
}\label{F18}}
\end{figure}

The purpose of some attacks is not decryption,
and the purpose is to destroy information, like
noise and data loss attacks.
In these attacks, parts of the information may be lost.
In the proposed algorithm, the location of pixels
are changed, and this can be a response to these attacks.
The simulation results for these attacks are given in Fig.s \ref{F17}-\ref{F18}.
In the Fig. \ref{F17}, the results show that
the proposed algorithm has better results than the algorithm in \cite{41}.
Also, in Fig. \ref{F18}, for different values of noise density, results are presented.
By using the results, it is seen that the encrypted images are recognizable.
Therefore, we can say that the proposed algorithm can resist noise and data
loss attacks.

\subsection{Simulation results of the steganography algorithm}
\begin{figure}
\centering
\includegraphics*[width=.80\textwidth]{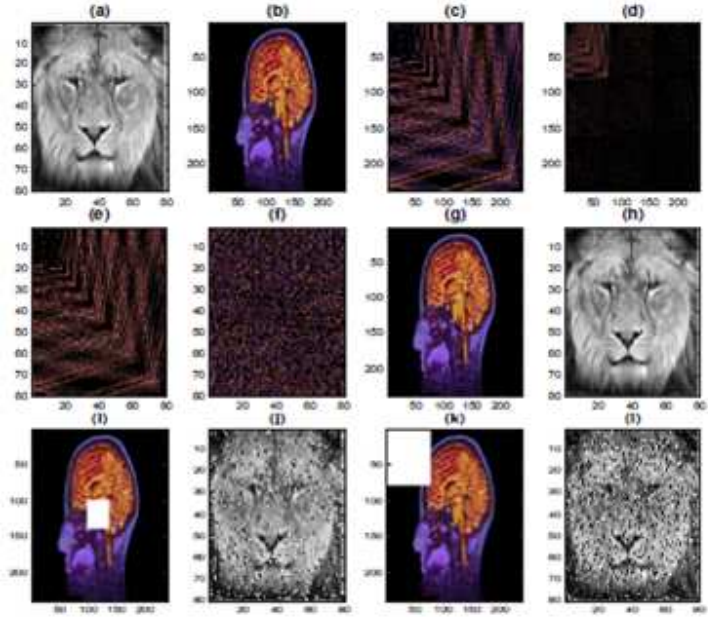}
\emph{\caption{
Results for steganography algorithm: (a) secret image,
(b) cover image, (c) output image of the step 3,
(d)-(f) output images of the step 4, (g) stego image,
(h) decrypted secret image,
(i),(k) data loss attack images, (j),
(l)decrypted secret images.
}\label{ST1}}
\end{figure}

In this section, the results of the proposed steganography algorithm
are given. In Fig. \ref{ST1}, we apply the proposed algorithm
on lion($80 \times 80$) and MRI($250 \times 250$) images as
secret and cover images, respectively. According to the proposed algorithm,
first of all, we shift the cover image by using two types of the spiral fashion
shifting (I and II)
after 1000 shifts (Fig. \ref{ST1}\textcolor{blue}{.c}).
Next, we use discrete framlete transform (DFT)(Fig. \ref{ST1}\textcolor{blue}{.d}).
Next, we cut the LL part of DFT (Fig. \ref{ST1}\textcolor{blue}{.e}).
Then in Fig. \ref{ST1}\textcolor{blue}{.f}, the odd and even columns are shifted up
and down and the odd
and even rows are shifted right and left by applying $x_0=0.1, y_0=0.3$ and $r=0.2$
as input values for
the proposed chaos map \big(for details, we refer to step4 (ii)\big). Note that
we can use the
proposed key generation function to get $x_0,y_0$ and $r$. Finally, we set the most
significant bits
of the secret image in the least significant
bits of the obtained image from the previous step (see step5).
By doing the above processes conversely, stego image (Fig. \ref{ST1}\textcolor{blue}{.g})
is obtained.
To show the resistance of the proposed algorithm against the data loss attacks,
the simulation results
for data loss attacks are given in Fig.s \ref{ST1}\textcolor{blue}{.i-l}, where
once the $40\times 40$
size of the central part   and once again in the size of the secret image,
i.e. $80\times 80$, of the corner
part of the stego image are removed.  \\

\begin{center}
\begin{table}[ht!]
\caption{ {\footnotesize PSNR values for steganography algorithm}}
\label{table:t2}
\centering
\begin{tabular}{ccccccccc}
\hline
&Fig. \ref{ST1}\textcolor{blue}{.g} &Fig. \ref{ST1}\textcolor{blue}{.h}&
Fig. \ref{ST1}\textcolor{blue}{.j}&Fig. \ref{ST1}\textcolor{blue}{.l}     \\
\hline\hline
PSNR&48.1594  &29.2125  &16.5766  &14.6422         \\
\hline
\end{tabular}
\end{table}
\end{center}

\newpage
\section{Appendix}
In this section, to make the visualization of the proposed steps easier, we present
some examples.\\

\begin{figure}
\centering
\includegraphics*[width=.80\textwidth]{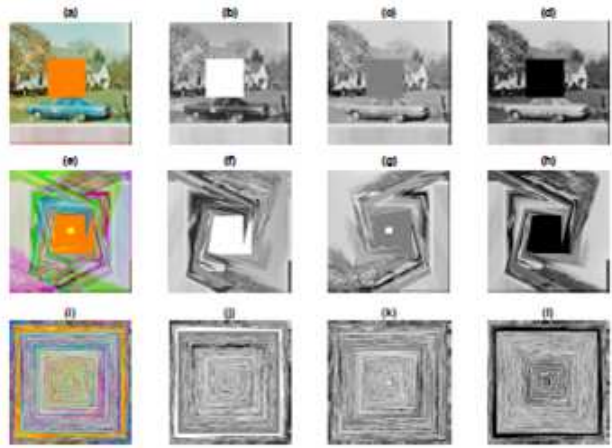}
\emph{\caption{
Showing the results of the mixing of an image by circular fashion shifting:
(a-d) original image, (e-h) circular fashion shifting after $190$ shifts
 (i-l) circular fashion shifting after $150000$ shifts.}\label{Ap2}}
\end{figure}

Fig. \ref{Ap2} represents the circular fashion shifting in the proposed encryption process.
For better
understanding, we apply this algorithm on an image of the size $256\times 256,$ that almost
the center of this changed to orange color.

\begin{figure}
\centering
\includegraphics*[width=.80\textwidth]{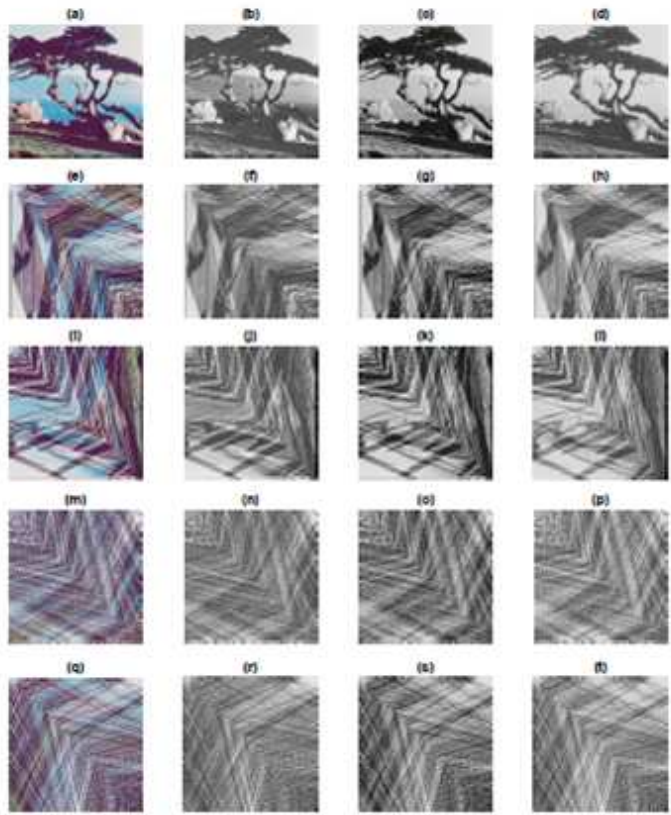}
\emph{\caption{
Showing the results of the mixing of the image by spiral fashion shifting:
(a-d) original image and its own component,  (e-h)  shifting by I type,
 (i-l) shifting by II type, (m-p)  shifting by I and II types, respectively,
 (q-t)  shifting by II and I types, respectively.}\label{Ap3}}
\end{figure}
\section{Conclusion}
In this paper, by introducing a 2D-hybride chaos map,
safe ways to transfer images securely by using
cryptography and steganography methods based on
 this chaos map are presented.
 Also,  in the proposed algorithm , framelet, cellular
automata and kinds of shifts have been used to increase
the resistance against different attacks.
By using simulation,
results, we can say that the proposed  algorithms can
 effectively resist differential,
statistical, noise, data loss and chosen-plain text attacks.

\end{document}